\documentclass[twocolumn,aps,superscriptaddress,showpacs,nofootinbib,floatfix]{revtex4-1}

\usepackage{multirow}
\usepackage{graphicx}
\usepackage{amsmath}

\usepackage{latexsym}
\usepackage{feynmp}

\usepackage{epsfig,bm,color}
\usepackage[normalem]{ulem}  % \sout{old text} for strikeout

\renewcommand\sout{\bgroup \color{red} \ULdepth=-.5ex \ULset}
\begin{document}

\title{The Doubly-heavy Tetraquarks ($qq'\bar{Q}\bar{Q'}$) in a Constituent Quark Model with a Complete Set of Harmonic Oscillator Bases}

\author{Sungsik Noh}
\email{sungsiknoh@yonsei.ac.kr}\affiliation{Department of Physics and Institute of Physics and Applied Physics, Yonsei
University, Seoul 03722, Korea}
\author{Woosung Park}
\email{diracdelta@yonsei.ac.kr}\affiliation{Department of Physics and Institute of Physics and Applied Physics, Yonsei
University, Seoul 03722, Korea}
\author{Su Houng Lee}
\email{suhoung@yonsei.ac.kr}\affiliation{Department of Physics and Institute of Physics and Applied Physics, Yonsei
University, Seoul 03722, Korea}

%\date{\today}
\begin{abstract}
We have improved our previous variational method based constituent quark model by introducing a complete set of 3-dimensional harmonic oscillator bases as the spatial part of the total wave function. To assess the validity of our approach, we compared the binding energy, thus calculated with the exact value for the hydrogen model.
After fitting to the masses of the ground state hadrons, we apply our new method to analyze the doubly-heavy tetraquark states $qq'\bar{Q}\bar{Q'}$ and compared the result for the binding energies with that from other works. We also calculated the ground state masses of $T_{sc} (ud\bar{s}\bar{c})$ and $T_{sb} (ud\bar{s}\bar{b})$ with $(I,S) = (0,1), (0,2)$. We found that $T_{bb} (ud\bar{b}\bar{b})$ and $us\bar{b}\bar{b}$, both with $(I,S) = (0,1)$, are stable against the two lowest threshold meson states with binding energies $-145$ MeV and $-42$ MeV, respectively. We further found that $T_{cb} (ud\bar{c}\bar{b})$ is near the lowest threshold. The spatial sizes for the tetraquarks are also discussed.
\end{abstract}

\maketitle

%%%%%%%%%%%%%%%%%%%%%%%%%%%%%%%%%%%%%%%%%%%%%%%%%%%%%%%%%%%%%%%%%%%%%%%%%%%%%%%%%%%%%%%%%%%%%%%
\section{INTRODUCTION}
%%%%%%%%%%%%%%%%%%%%%%%%%%%%%%%%%%%%%%%%%%%%%%%%%%%%%%%%%%%%%%%%%%%%%%%%%%%%%%%%%%%%%%%%%%%%%%%
Since the observation of X(3872)\cite{Choi:2003ue} and several exotic hadron candidates that followed, the structure of these particles and other potential flavor exotic configurations have become a central theme of study.  Theoretical approaches on these topics range from direct lattice calculation\cite{Ikeda:2016zwx}, sum rule approach\cite{Nielsen:2009uh}, effective models and constituent quark based models\cite{Liu:2019zoy}. To better understand the data and to point to future searches, it becomes necessary to improve any simple minded model calculations to full details.    

Quark model based on color and spin interaction was successful in describing the mass differences between hadrons within a multiplet\cite{DeRujula:1975qlm}.
A constituent quark model with a more realistic potential was proposed in Ref.~\cite{Bhaduri} in the early 1980s, which gives a unified description of meson and baryon spectra. The potential in BCN model\cite{Bhaduri} was applied to investigate the baryon spectra in detail\cite{Silvestre:PRD1985}, and to various tetraquark states\cite{Silvestre:ZPC1993}. In Ref.~\cite{Silvestre:ZPC1994}, the authors introduced four different types of interquark potentials, which improves the simultaneous fit to both the meson and the baryon spectra.

In addition to building up the realistic potentials, there have been many works towards improving the accuracy of the spatial part of the total wave function. Among them, in Refs.~\cite{Silvestre:ZPC1986, Silvestre:ZPC1993}, the harmonic oscillator bases were applied to construct the spatial part of the wave function for the baryon structure. In particular, Ref.~\cite{Silvestre:ZPC1993} discussed the validity of the harmonic oscillator bases within the light baryons, composed of only $u$, $d$, and $s$ quarks. They also extended their work to the tetraquarks in Refs.~\cite{Silvestre:ZPC1986, Silvestre:ZPC1993,Silvestre:ZPC1994}. In Refs.~\cite{Silvestre:ZPC1986,Vijande:PRD2009}, the authors built the spatial function using the hyper spherical coordinates for the tetraquark systems. In Ref.~\cite{Brink:1998}, using the spatial function that is made up of multiple Gaussians which allow for internal angular momentum and satisfy the permutation symmetries restricted by the Pauli principle, the authors studied the properties of  $T_{cc}$ and $T_{bb}$, and confirmed the stability for the latter. Subsequently, in Ref.~\cite{Janc:FewBody2004}, the authors performed a similar calculations but with a more sophisticated bases of multiple Gaussians.

In this work, we have introduced a complete set of 3-dimensional harmonic oscillator bases with a rescaling factor that can be flexibly used for better convergence compared to the harmonic oscillator bases in the work\cite{Silvestre:ZPC1986}, and applied them to construct the spatial part of the total wave function in a constituent quark model with hyperfine potential given in Eq.~(\ref{Hamiltonian}).
In doing so, our spatial bases more efficiently describes the ground state wave functions not only for the meson and baryon structures, but also for the tetraquark structure.

In the following section, we first introduce the constituent quark model and compare the fittings in the present work to those in our previous work\cite{Woosung:NPA2019}.
In Section III, each part of the total wave function is introduced, emphasizing the classificiation of the harmonic oscillator bases with the notion of quanta of the harmonic oscillator bases.
In Section IV, the newly obtained numerical results are presented and compared with those in the previous work\cite{Woosung:NPA2019}. Furthermore, we show the relative distances between the quark pairs together with a pictorial description of the spatial sizes for the tetraquarks. We also discuss the tetraquarks as a baryon structure. Discussion and summary are given in Section V.
In Appendix, to assess the validity of our approach, we compare the result obtained with our method for the hydrogen model with the exact solution.
We also show the details in constructing the spatial wave function with the complete set of harmonic oscillator bases for the meson as well as the baryon structure. As a special case, in Appendix C, we present the method for constructing the bases for the proton.

%%%%%%%%%%%%%%%%%%%%%%%%%%%%%%%%%%%%%%%%%%%%%%%%%%%%%%%%%%%%%%%%%%%%%%%%%%%%%%%%%%%%%%%%%%%%%%%
\section{FORMALISM}
%%%%%%%%%%%%%%%%%%%%%%%%%%%%%%%%%%%%%%%%%%%%%%%%%%%%%%%%%%%%%%%%%%%%%%%%%%%%%%%%%%%%%%%%%%%%%%%
{}\

We use a nonrelativistic Hamiltonian for the constituent quarks, which is the same as in our previous work\cite{Woosung:NPA2019}.
\begin{eqnarray}
H &=& \sum^{4}_{i=4} \left( m_i+\frac{{\mathbf p}^{2}_i}{2 m_i} \right)-\frac{3}{4}\sum^{4}_{i<j}\frac{\lambda^{c}_{i}}{2} \,\, \frac{\lambda^{c}_{j}}{2} \left( V^{C}_{ij} + V^{CS}_{ij} \right), \qquad
\label{Hamiltonian}
\end{eqnarray}
where $m_i$ is the quark mass and $\lambda^c_{i}/2$ is the SU(3) color operator for the $i$-th quark. The internal quark potentials $V^{C}_{ij}$ and $V^{CS}_{ij}$ are taken to be the same forms as in our previous work\cite{Woosung:NPA2019}: 
\begin{eqnarray}
V^{C}_{ij} &=& - \frac{\kappa}{r_{ij}} + \frac{r_{ij}}{a^2_0} - D,
\label{ConfineP}
\\
V^{CS}_{ij} &=& \frac{\hbar^2 c^2 \kappa'}{m_i m_j c^4} \frac{e^{- \left( r_{ij} \right)^2 / \left( r_{0ij} \right)^2}}{(r_{0ij}) r_{ij}} \boldsymbol{\sigma}_i \cdot \boldsymbol{\sigma}_j.
\label{CSP}
\end{eqnarray}
Here
\begin{eqnarray}
r_{0ij} &=& 1/ \left( \alpha + \beta \frac{m_i m_j}{m_i + m_j} \right)	,	
\label{Parameter1}
\\
\kappa' &=& \kappa_0 \left( 1+ \gamma \frac{m_i m_j}{m_i + m_j} \right)	,	
\label{Parameter2}
\end{eqnarray}
where $r_{ij}=|{\mathbf r}_i - {\mathbf r}_j |$ is the relative distance between the $i$ and $j$ quarks, and $\boldsymbol{\sigma}_i$ is the spin operator. The parameters appearing in Eqs.~(\ref{ConfineP})-(\ref{Parameter2}) are determined by fitting them to the experimental ground state masses of the hadrons listed in Tables~\ref{mesons}, \ref{baryons}.
%%%%%%%%%%%%%%%%%%%%%%%%%%%		Table 1. Meson masses		%%%%%%%%%%%%%%%%%%%%%%%%%%%%%%%%%%%%%%
\begin{table}[t]

\caption{The masses of mesons obtained (Column 3) with the fitting parameters set given in  Eq.~(\ref{FitParameters}).  Column 4 shows the variational parameter $a$ .}	

\centering

\begin{tabular}{ccccc}
\hline
\hline	\multirow{2}{*}{Particle}	&	Experimental	&	Mass	&	Variational	&	Error	\\
										&	Value (MeV)	&	(MeV)	&	Parameter (${\rm fm}^{-2}$)	&	(\%)	\\
\hline 
$D$			&	1864.8		&	1853.8		&	$a$ = 7.5		&	0.59	\\
$D^*$		&	2007.0		&	2006.2		&	$a$ = 5.7		&	0.04	\\
$\eta_{c}$	&	2983.6		&	2986.0		&	$a$ = 25.2		&	0.08	\\
$J/\Psi$	&	3096.9		&	3118.4		&	$a$ = 19.7		&	0.69	\\
$D_s$		&	1968.3		&	1963.6		&	$a$ = 12.1		&	0.24	\\
$D^*_s$		&	2112.1		&	2109.2		&	$a$ = 9.3		&	0.14	\\
$K$			&	493.68		&	498.32		&	$a$ = 7.7		&	0.94	\\
$K^*$		&	891.66	 	&	874.66		&	$a$ = 4.1		&	1.91	\\

$B$			&	5279.3	 	&	5301.2		&	$a$ = 7.3		&	0.42	\\
$B^*$		&	5325.2		&	5360.5		&	$a$ = 6.5		&	0.66	\\
$\eta_b$	&	9398.0		&	9327.1		&	$a$ = 100.2	&	0.75	\\
$\Upsilon$	&	9460.3		&	9456.6		&	$a$ = 81.9		&	0.04	\\
$B_s$		&	5366.8		&	5375.3		&	$a$ = 13.0		&	0.16	\\
$B_s^*$		&	5415.4		&	5439.3		&	$a$ = 11.5		&	0.44	\\
$B_c$		&	6275.6		&	6268.4		&	$a$ = 38.7		&	0.11	\\
$B_c^*$		&		-		&	6361.8		&	$a$ = 32.6		&	-		\\
\hline 
\hline
\label{mesons}
\end{tabular}
\end{table}
%%%%%%%%%%%%%%%%%%%%%%%%%%%		Table 2. Baryon masses		%%%%%%%%%%%%%%%%%%%%%%%%%%%%%%%%
\begin{table}[t]
\caption{Same as Table~\ref{mesons} but for baryons. In column 4, $a_1$ and $a_2$ are the variational parameters.}

\centering

\begin{tabular}{ccccc}
\hline
\hline	\multirow{2}{*}{Particle}	&	Experimental	&	Mass	&	\quad Variational\quad	&	Error	\\
										&	Value (MeV)	&	(MeV)	&	\quad Parameters (${\rm fm}^{-2}$)\quad	&	(\%)	\\
\hline  
$\Lambda$		&	1115.7	&	1111.7	&	\quad$a_1$ = 4.2, $a_2$ = 3.5\quad	&	0.36	\\
$\Lambda_{c}$	&	2286.5	&	2269.4	&	\quad$a_1$ = 4.4, $a_2$ = 4.4\quad	&	0.75	\\
$\Xi_{cc}$		&	3621.4	&	3621.1	&	\quad$a_1$ = 11.3, $a_2$ = 4.5\quad	&	0.01	\\
$\Lambda_b$	&	5619.4	&	5634.6	&	\quad$a_1$ = 4.5, $a_2$ = 5.0\quad	&	0.27	\\
$\Sigma_{c}$	&	2452.9	&	2438.5	&	\quad$a_1$ = 2.8, $a_2$ = 5.5\quad	&	0.59	\\
$\Sigma_{c}^*$	&	2517.5	&	2523.2	&	\quad$a_1$ = 2.5, $a_2$ = 4.7\quad	&	0.23	\\
$\Sigma_{b}$	&	5811.3	&	5841.6	&	\quad$a_1$ = 2.8, $a_2$ = 5.8\quad	&	0.52	\\
$\Sigma_{b}^*$	&	5832.1	&	5875.1	&	\quad$a_1$ = 2.8, $a_2$ = 4.2\quad	&	0.74	\\
$\Sigma$		&	1192.6	&	1192.4	&	\quad$a_1$ = 2.9, $a_2$ = 4.6\quad	&	0.02	\\
$\Sigma^*$		&	1383.7	&	1395.8	&	\quad$a_1$ = 2.3, $a_2$ = 3.2\quad	&	0.87	\\
$\Xi$			&	1314.9	&	1327.1	&	\quad$a_1$ = 4.5, $a_2$ = 4.3\quad	&	0.93	\\
$\Xi^*$			&	1531.8	&	1540.4	&	\quad$a_1$ = 4.1, $a_2$ = 2.8\quad	&	0.56	\\
$\Xi_{c}$		&	2467.8	&	2472.0	&	\quad$a_1$ = 4.9, $a_2$ = 6.0\quad	&	0.17	\\
$\Xi_{c}^*$	&	2645.9	&	2649.7	&	\quad$a_1$ = 3.3, $a_2$ = 6.1\quad	&	0.15	\\
$\Xi_{b}$		&	5787.8	&	5824.1	&	\quad$a_1$ = 5.0, $a_2$ = 7.2\quad	&	0.63	\\
$\Xi_{b}^*$	&	5945.5	&	5989.1	&	\quad$a_1$ = 3.4, $a_2$ = 7.6\quad	&	0.73	\\

$p$				&	938.27	&	938.05	&	\quad$a_1$ = 2.9, $a_2$ = 2.9\quad	&	0.02	\\
$\Delta$		&	1232	&	1242.2	&	\quad$a_1$ = 1.9, $a_2$ = 1.9\quad	&	0.83	\\
\hline 
\hline
\label{baryons}
\end{tabular}
\end{table}
%%%%%%%%%%%%%%%%%%%%%%%%%%%%%%%%%%%%%%%%%%%%%%%%%%%%%%%%%%%%%%%%%%%%%%%%%%%%%%%%%%%%%%%%%%%%%

The fitting has been done mostly within 0.7\% error as can be seen in Tables~\ref{mesons}, \ref{baryons}. In particular, we focused on trying to minimize the error on the doubly-charmed baryon, $\Xi_{cc}$, and on the baryons $\Lambda$, $p$ and also on the mesons $D$, $D^*$, $B$, $B^*$, $B_s$ which comprise the lowest thresholds for the doubly-heavy tetraquarks of interest in this work.

In comparison to our previous work\cite{Woosung:NPA2019}, where only hadrons with at least one $c$ or $b$ were fit, here we fit all hadrons including those with only light quarks. The number of fitting hadrons is increased by almost a factor of 2 compared to that in our previous work\cite{Woosung:NPA2019}. 
The better fit is a consequence of using the complete set of harmonic oscillator bases approach.
Using the harmonic oscillator bases, the new fitting parameters are as follows.
\\
\begin{eqnarray}
&\kappa=120.0 \, \textrm{MeV fm}, \quad a_0=0.0334066 \, \textrm{(MeV$^{-1}$fm)$^{1/2}$},& \nonumber \\
&D=917  \, \textrm{MeV}, & \nonumber \\
&m_{u}=342 \, \textrm{MeV}, \qquad m_{s}=642 \, \textrm{MeV}, &\nonumber \\
&m_{c}=1922 \, \textrm{MeV}, \qquad m_{b}=5337 \, \textrm{MeV},	&\nonumber \\
&\alpha = 1.0749 \, \textrm{fm$^{-1}$}, \,\, \beta = 0.0008014 \, \textrm{(MeV fm)$^{-1}$}, &	\nonumber \\
&\gamma = 0.001380 \, \textrm{MeV$^{-1}$}, \,\, \kappa_0=197.144 \, \textrm{MeV}. 	 &
\label{FitParameters}
\end{eqnarray}
\\

%%%%%%%%%%%%%%%%%%%%%%%%%%%%%%%%%%%%%%%%%%%%%%%%%%%%%%%%%%%%%%%%%%%%%%%%%%%%%%%%%%%%%%%%%%%%%%%
\section{Wave Function}
%%%%%%%%%%%%%%%%%%%%%%%%%%%%%%%%%%%%%%%%%%%%%%%%%%%%%%%%%%%%%%%%%%%%%%%%%%%%%%%%%%%%%%%%%%%%%%%
Here, we will study the tetraquark states with the total orbital angular momentum equal to zero in the constituent quark model.   
However, it should be noted that the $p$-wave or even higher orbital states between the quarks can play a crucial role in lowering the total hadron energy and contribute to the total wave function. This can be successfully done by introducing the complete set of 3-dimensional harmonic oscillator bases. The total wave function of the Hamiltonian consists of the spatial, color, spin, and the flavor parts of bases. We adopt the harmonic oscillator bases as the spatial part of the wave function. The other parts of the wave function are the same as in the work\cite{Woosung:NPA2019}. We will thus discuss the spatial part in detail, while the other parts will be mentioned briefly.

%%%%%%%%%%%%%%%%%%%%%%%%%%%%%			Section III A			%%%%%%%%%%%%%%%%%%%%%%%%%%%%%%%%%%
\subsection{Jacobi Coordinates Sets}
%%%%%%%%%%%%%%%%%%%%%%%%%%%%%%%%%%%%%%%%%%%%%%%%%%%%%%%%%%%%%%%%%%%%%%%%%%%%%%%%%%%%%%%%%%%%%%%
To set up the spatial function, we first set the Jacobi coordinates, representing the relative positions of all the quarks in the tetraquark configuration. The Jacobi coordinates sets in each configuration can be written as follows.
\begin{itemize}
	\item{Coordinates Set 1}
	\begin{eqnarray}
	& \mathbf{x}_1 = \frac{1}{\sqrt{2}}({\mathbf r}_1 - {\mathbf r}_2), \qquad \mathbf{x}_2 = \frac{1}{\sqrt{2}}({\mathbf r}_3 - {\mathbf r}_4)\,, &	\nonumber
	\\
	& \mathbf{x}_3 = \frac{1}{\mu} \left( \frac{m_1 {\mathbf r}_1 + m_2 {\mathbf r}_2}{m_1 + m_2} - \frac{m_3 {\mathbf r}_3 + m_4 {\mathbf r}_4}{m_3 + m_4} \right)\,, &
\label{CoordSet1}
	\end{eqnarray}
	\\
	\item{Coordinates Set 2}
	\begin{eqnarray}
	& \mathbf{y}_1 = \frac{1}{\sqrt{2}}({\mathbf r}_1 - {\mathbf r}_3), \qquad \mathbf{y}_2 = \frac{1}{\sqrt{2}}({\mathbf r}_4 - {\mathbf r}_2)\,, &	\nonumber
	\\
	& \mathbf{y}_3 = \frac{1}{\mu} \left( \frac{m_1 {\mathbf r}_1 + m_2 {\mathbf r}_3}{m_1 + m_2} - \frac{m_3 {\mathbf r}_2 + m_4 {\mathbf r}_4}{m_3 + m_4} \right)\,, &
	\end{eqnarray}
	\\
	\item{Coordinates Set 3}
	\begin{eqnarray}
	& \mathbf{z}_1 = \frac{1}{\sqrt{2}}({\mathbf r}_1 - {\mathbf r}_4), \qquad \mathbf{z}_2 = \frac{1}{\sqrt{2}}({\mathbf r}_2 - {\mathbf r}_3)\,, &	\nonumber
	\\
	& \mathbf{z}_3 = \frac{1}{\mu} \left( \frac{m_1 {\mathbf r}_1 + m_2 {\mathbf r}_4}{m_1 + m_2} - \frac{m_3 {\mathbf r}_2 + m_4 {\mathbf r}_3}{m_3 + m_4} \right)\,, &
	\end{eqnarray}
\end{itemize}
where
\begin{eqnarray}
\mu &=& \left[ \frac{m_1^2 + m_2^2}{(m_1 + m_2)^2} + \frac{m_3^2 + m_4^2}{(m_3 + m_4)^2} \right]^{1/2} \,,	\nonumber
\end{eqnarray}
and
\begin{eqnarray}
&m_u=m_d;&	\nonumber
\\
&m_1=m_2=m_u,	\,\,	m_3=m_4=m_c& 	\quad {\rm for} \,\, ud\bar{c}\bar{c} ,	\nonumber
\\
&m_1=m_2=m_u,	\,\,	m_3=m_4=m_b& 	\quad {\rm for} \,\, ud\bar{b}\bar{b},	\nonumber
\\
&m_1=m_2=m_u,	\,\,	m_3=m_c, \,\, m_4=m_b& 	\quad {\rm for} \,\, ud\bar{c}\bar{b},	\nonumber
\\
&m_1=m_2=m_u,	\,\,	m_3=m_s, \,\, m_4=m_c& 	\quad {\rm for} \,\, ud\bar{s}\bar{c},	\nonumber
\\
&m_1=m_2=m_u,	\,\,	m_3=m_s, \,\, m_4=m_b& 	\quad {\rm for} \,\, ud\bar{s}\bar{b},	\nonumber
\\
&m_1=m_u, \,\, m_2=m_s,	\,\,	m_3=m_4=m_b& 	\quad {\rm for} \,\, us\bar{b}\bar{b}. \nonumber
\end{eqnarray}
For symmetry reason, we take the coordinates set 1 as our reference, and use the transformations between the above sets of Jacobi coordinates to calculate relevant matrix elements involving two quarks.

%%%%%%%%%%%%%%%%%%%%%%%%%%%%%			Section III B			%%%%%%%%%%%%%%%%%%%%%%%%%%%%%%%%%%
\subsection{Color, Spin Bases and Flavor}
%%%%%%%%%%%%%%%%%%%%%%%%%%%%%%%%%%%%%%%%%%%%%%%%%%%%%%%%%%%%%%%%%%%%%%%%%%%%%%%%%%%%%%%%%%%%%%%
The most stable state for the doubly-heavy tetraquarks can be found in the spin 1 channel\cite{Woosung:NPA2019}. The colosr-spin (CS) space for the spin 1 tetraquark system is spanned by the six CS bases due to the fact that the tetraquark configuration in the total spin 1 state can be described by two color bases, and three spin bases. In the configuration of the Jacobi coordinates set 1, the six CS bases are as follows\cite{Woosung:NPA2019}.
\begin{eqnarray}
\hspace{-0.6cm}
&\psi^{CS}_1 = \left( q_1 q_2 \right)^{\mathbf{6}}_{1} \otimes \left( \bar{q}_3 \bar{q}_4 \right)^{\mathbf{\bar{6}}}_{1}
, \,\,
\psi^{CS}_2 = \left( q_1 q_2 \right)^{\mathbf{\bar{3}}}_{1} \otimes \left( \bar{q}_3 \bar{q}_4 \right)^{\mathbf{3}}_{1} , &
\nonumber \\
\hspace{-0.6cm}
&\psi^{CS}_3 = \left( q_1 q_2 \right)^{\mathbf{6}}_{1} \otimes \left( \bar{q}_3 \bar{q}_4 \right)^{\mathbf{\bar{6}}}_{0}
, \,\,
\psi^{CS}_4 = \left( q_1 q_2 \right)^{\mathbf{\bar{3}}}_{1} \otimes \left( \bar{q}_3 \bar{q}_4 \right)^{\mathbf{3}}_{0} , &
\nonumber \\
\hspace{-0.6cm}
&\psi^{CS}_5 = \left( q_1 q_2 \right)^{\mathbf{6}}_{0} \otimes \left( \bar{q}_3 \bar{q}_4 \right)^{\mathbf{\bar{6}}}_{1}
, \,\,
\psi^{CS}_6 = \left( q_1 q_2 \right)^{\mathbf{\bar{3}}}_{0} \otimes \left( \bar{q}_3 \bar{q}_4 \right)^{\mathbf{3}}_{1}. &
\end{eqnarray}
where the superscript indicates the color state, and the subscript indicates the spin state for the subparticle systems in the tetraquark structure.

For the falvor part, we will consider the isospin 0 for $T_{cc} (ud\bar{c}\bar{c})$, $T_{bb} (ud\bar{b}\bar{b})$, $T_{cb} (ud\bar{c}\bar{b})$, $T_{sc} (ud\bar{s}\bar{c})$, $T_{sb} (ud\bar{s}\bar{b})$, and the isospin 1/2 for $us\bar{b}\bar{b}$. The basis of the Hamiltonian is determined to satisfy the symmetry constraint due to the Pauli principle, and constructed by combining the CS basis with the spatial part. The permutation symmetries for the CS bases and for the flavor part are summarized in Table~\ref{SymmetryCSF}.

%%%%%%%%%%%%%%%%%%%%%		Table 3. Symmetries of CS and Flavor		%%%%%%%%%%%%%%%%%%%%%%%%%
\begin{table}[t]

\caption{The permutation symmetries of the color, spin, and flavor parts for the tetraquarks studied in this work.   Here, $+1(-1)$ indicates that it is symmetric (antisymmetric) under the corresponding permutations in the configuration of the Jacobi coordinates set 1. The blank in the table indicates that there is no symmetry constraint under the corresponding permutation. $T_{QQ'}$ stands for $ud\bar{Q}\bar{Q'}$.}	
\centering
\begin{tabular}{cccccccc}
\hline
\hline
\multirow{2}{*}{Permutation}	&&	\multicolumn{6}{c}{CS Bases}		\\
\cline{3-8}
	&&	$\psi^{CS}_1$	&	$\psi^{CS}_2$	&	$\psi^{CS}_3$	&	$\psi^{CS}_4$	&	$\psi^{CS}_5$	&	$\psi^{CS}_6$	\\
\hline
(12)	&&	+1	&	$-1$&	+1	&	$-1$&$-1$	&	+1	\\
(34)	&&	+1	&	$-1$&	$-1$&	+1	&	+1	&	$-1$\\
\hline
\hline
\multirow{2}{*}{Permutation}	&&	\multicolumn{6}{c}{Flavor}	\\
\cline{3-8}
	&&	\multicolumn{2}{c}{$T_{QQ}$}	&	\multicolumn{2}{c}{$T_{QQ'}$}	&	\multicolumn{2}{c}{$us\bar{b}\bar{b}$}	\\
\hline
$(12)$		&&	\multicolumn{2}{c}{$-1$}	&	\multicolumn{2}{c}{$-1$}	&	\multicolumn{2}{c}{}	\\
$(34)$		&&	\multicolumn{2}{c}{$+1$}	&	\multicolumn{2}{c}{}	&	\multicolumn{2}{c}{+1}	\\
\hline 
\hline
\label{SymmetryCSF}
\end{tabular}
\end{table}
%%%%%%%%%%%%%%%%%%%%%%%%%%%%%%%%%%%%%%%%%%%%%%%%%%%%%%%%%%%%%%%%%%%%%%%%%%%%%%%%%%%%%%%%%%%%%%

%%%%%%%%%%%%%%%%%%%%%%%%%%%%%			Section III C				%%%%%%%%%%%%%%%%%%%%%%%%%%%%%%%
\subsection{Harmonic Oscillator Bases as The Spatial Function}
\label{Section3C}
%%%%%%%%%%%%%%%%%%%%%%%%%%%%%%%%%%%%%%%%%%%%%%%%%%%%%%%%%%%%%%%%%%%%%%%%%%%%%%%%%%%%%%%%%%%%%%%
Generalizing the method used in Appendix A, and B for the meson and baryon structures, we construct the complete set of harmonic oscillator bases for the tetraquarks. Since there are three internal orbital angular momenta $l_1$, $l_2$, and $l_3$, there are three ways to  combine them depending on the order of addition. Choosing to combine $l_1$ and $l_2$ first, the spatial function can be constructed as follows.
\begin{eqnarray}
&&\hspace{-0.5cm}\psi( \mathbf{x}_1, \mathbf{x}_2, \mathbf{x}_3)^{Spatial}_{[n_1,n_2,n_3,l_1,l_2,l_3]}
\nonumber \\
&&\hspace{-0.3cm}= \sum_{m_1,m_2,m_3} C(l_1,m_1,l_2,m_2;L_{1,2}=l_3,m_{1,2}=-m_3)
\nonumber\\
&&\hspace{0.1cm} \times \, C(L_{1,2}=l_3,m_{1,2}=-m_3,l_3,m_3;l=0,m=0)
\nonumber \\
&&\hspace{0.1cm} \times \, R_{n_1,l_1}(x_1) R_{n_2,l_2}(x_2) R_{n_3,l_3}(x_3)
\nonumber \\
&&\hspace{0.1cm} \times \, Y^{m_1}_{l_1} (\theta_1, \phi_1) Y^{m_2}_{l_2} (\theta_2, \phi_2) Y^{m_3}_{l_3} (\theta_3, \phi_3) \,.
\label{TetraSpatial}
\end{eqnarray}
where the coefficient $C(l_1,m_1,l_2,m_2;L_{1,2},m_{1,2})$ is the Clebsch-Gordan(CG) coefficient for the decomposition of the subtotal angular state $|L_{1,2}, m_{1,2}\rangle$ in terms of $|l_1, m_1\rangle |l_2, m_2\rangle$, and the subtotal angular state of the subparticle corresponding to the quark pair $(1,2)$ is restricted to $|L_{1,2}=l_3, m_{1,2}=-m_3\rangle$ to satisfy the total $l=0$ state. $R_{n_i,l_i}(x_i)$ has the same form as in Eq.~(\ref{MesonSpatial}), and $Y^{m_i}_{l_i} (\theta_i, \phi_i)$ is the spherical harmonic function for the angular part of the $i$-th Jacobi coordinate $\mathbf{x}_i$. For the other two types of combinations, Eq.~(\ref{TetraSpatial}) is modified by adding $l_1,l_3$ or $l_2,l_3$ first.

Since there are a large number of harmonic oscillator bases in the tetraquark system, there has to be a criterion when adding the harmonic oscillator bases as a spatial function into the calculations. The magnitude of the eigenvalues of the Hamiltonian is mostly affected by the diagonal elements rather than the off-diagonal ones. Especially, except for the rest masses, the largest part in the diagonal element is the kinetic energy. In addition, non-zero kinetic energy contribution is mostly coming from the diagonal elements. Therefore, in classifying the harmonic oscillator bases, we define the quanta of the harmonic oscillator bases by the expectation value of the kientic energy of the diagonal component for the extreme case where all the constituent quark masses are identical. Introducing the center of mass frame, the kinetic energy denoted by $T_c$ becomes as follows.
\begin{eqnarray}
T_c
&=& \sum^4_{i=1} \frac{{\mathbf p}^2_i}{2m_i} - \frac{{\mathbf p}^2_{rC}}{2M}
=\frac{{\mathbf p}^2_{1}}{2m'_1} + \frac{{\mathbf p}^2_{2}}{2m'_2} + \frac{{\mathbf p}^2_{3}}{2m'_3} \,,
\end{eqnarray}
where
\begin{eqnarray}
&&m'_1=m_u, \,\, m'_2=m_c, \,\, m'_3=\frac{2 m_u m_c}{m_u + m_c}	\quad\qquad {\rm for} \,\, ud\bar{c}\bar{c},
\nonumber \\
&&m'_1=m_u, \,\, m'_2=m_b, \,\, m'_3=\frac{2 m_u m_b}{m_u + m_b}	\quad\qquad {\rm for} \,\, ud\bar{b}\bar{b},
\nonumber \\
&&m'_1=m_u, \,\, m'_2=\frac{2 m_c m_b}{m_c + m_b}, \nonumber \\
&&\hspace{0.28cm} m'_3=\frac{(3 m_c^2 + 2 m_c m_b + 3 m_b^2) m_u}{(m_c + m_b) (2 m_u + m_c + m_b)}  \,\,\quad\qquad {\rm for} \,\, ud\bar{c}\bar{b},
\nonumber \\
&&m'_1=m_u, \,\, m'_2=\frac{2 m_s m_c}{m_s + m_c}, 	\nonumber \\
&&\hspace{0.28cm} m'_3=\frac{(3 m_s^2 + 2 m_s m_c + 3 m_c^2) m_u}{(m_s + m_c) (2 m_u + m_s + m_c)}  \,\,\quad\qquad {\rm for} \,\, ud\bar{s}\bar{c},
\nonumber \\
&&m'_1=m_u, \,\, m'_2=\frac{2 m_s m_b}{m_s + m_b}, 	\nonumber \\
&&\hspace{0.28cm} m'_3=\frac{(3 m_s^2 + 2 m_s m_b + 3 m_b^2) m_u}{(m_s + m_b) (2 m_u + m_s + m_b)}  \,\,\quad\qquad {\rm for} \,\, ud\bar{s}\bar{b},
\nonumber	\\
&&m'_1=\frac{2 m_u m_s}{m_u + m_s}, \,\, m'_2=m_b, 	\nonumber \\
&&\hspace{0.28cm} m'_3=\frac{(3 m_u^2 + 2 m_u m_s + 3 m_s^2) m_b}{(m_u + m_s) (m_u + m_s + 2 m_b)}  \,\,\quad\qquad {\rm for} \,\, us\bar{b}\bar{b}.	\nonumber
\\
\label{ReducedM}
\end{eqnarray}

For the extreme case where all the masses of the constituent quarks are identical ($m_1=m_2=m_3=m_4 \equiv m$), and thus also the variational parameters ($a_1=a_2=a_3 \equiv a$), the diagonal component of the kinetic energy reduces to
\begin{small}
\begin{eqnarray}
\langle T_c \rangle
&=&
\frac{\hbar^2 c^2 a}{m} \bigg[ \left(2 n_1 + l_1 + \frac{3}{2} \right)
\nonumber \\
&& \qquad \qquad \qquad + \left(2 n_2 + l_2 + \frac{3}{2} \right) + \left(2 n_3 + l_3 + \frac{3}{2} \right) \bigg]
\nonumber \\
&=&
\frac{\hbar^2 c^2 a}{m} \bigg[ 2 n_1 + l_1 + 2 n_2 + l_2 + 2 n_3 + l_3 + \frac{9}{2} \bigg] \,.
\label{kineticQuanta}
\end{eqnarray}
\end{small}
\hspace{-0.15cm}Therefore, one notes for this special cases that the kinetic energy is the same for all the possible combinations of the quantum numbers ($n_1, n_2, n_3, l_1, l_2, l_3$) if the sum $(Q=2 n_1  + 2 n_2 + 2 n_3 + l_1 + l_2 + l_3 )$ is unchanged. One can now organize the harmonic oscillator bases according to the quanta $Q$ appearing in Eq.~(\ref{kineticQuanta}). Then, for mesons, each of the spatial functions is categorized in the different quanta. For the tetraquarks, the spatial functions included in our calculations are classified as follows. In terms of $\psi^{spatial}_{[n_1, n_2, n_3, l_1, l_2, l_3]}$ without specifying the arguments $( \mathbf{x}_1,  \mathbf{x}_2,  \mathbf{x}_3)$,
%%%%%%%%%%%%%%%%%%%%%%%%%%%%%%%%%%%%%%%%%%%%%%%%%%%%%%%%%%%%%%%%%%%%%%%%%%%%%%%%%%%%%%%%%
%
%			Tetraquark Spatial Functions used in calculations
%
%%%%%%%%%%%%%%%%%%%%%%%%%%%%%%%%%%%%%%%%%%%%%%%%%%%%%%%%%%%%%%%%%%%%%%%%%%%%%%%%%%%%%%%%%
\begin{itemize}
%%%%%%%%%%%%%%%%%%%%%%%%%%%%%%%%%%%%%%%%%%
	\begin{item}1st Quanta ($Q=0$) \\
	$\psi^{Spatial}_{[0,0,0,0,0,0]}$.
	\end{item}
	\\
%%%%%%%%%%%%%%%%%%%%%%%%%%%%%%%%%%%%%%%%%%
	\begin{item}2nd Quanta ($Q=2$)\\
	$\psi^{Spatial}_{[1,0,0,0,0,0]}$, $\psi^{Spatial}_{[0,1,0,0,0,0]}$, $\psi^{Spatial}_{[0,0,1,0,0,0]}$,
	\\
	$\psi^{Spatial}_{[0,0,0,1,1,0]}$, $\psi^{Spatial}_{[0,0,0,1,0,1]}$, $\psi^{Spatial}_{[0,0,0,0,1,1]}$.
	\end{item}
	\\
%%%%%%%%%%%%%%%%%%%%%%%%%%%%%%%%%%%%%%%%%%
	\begin{item}3rd Quanta ($Q=4$)\\
	$\psi^{Spatial}_{[2,0,0,0,0,0]}$, $\psi^{Spatial}_{[0,2,0,0,0,0]}$, $\psi^{Spatial}_{[0,0,2,0,0,0]}$,
	\\
	$\psi^{Spatial}_{[1,0,0,1,1,0]}$, $\psi^{Spatial}_{[0,1,0,1,1,0]}$, $\psi^{Spatial}_{[0,0,1,1,1,0]}$,
	\\
	$\psi^{Spatial}_{[1,0,0,1,0,1]}$, $\psi^{Spatial}_{[0,1,0,1,0,1]}$, $\psi^{Spatial}_{[0,0,1,1,0,1]}$,
	\\
	$\psi^{Spatial}_{[1,0,0,0,1,1]}$, $\psi^{Spatial}_{[0,1,0,0,1,1]}$, $\psi^{Spatial}_{[0,0,1,0,1,1]}$,
	\\
	$\psi^{Spatial}_{[0,0,0,2,2,0]}$, $\psi^{Spatial}_{[0,0,0,2,0,2]}$, $\psi^{Spatial}_{[0,0,0,0,2,2]}$,
	\\
	$\psi^{Spatial}_{[0,0,0,1,1,2]}$, $\psi^{Spatial}_{[0,0,0,1,2,1]}$, $\psi^{Spatial}_{[0,0,0,2,1,1]}$,
	\\
	$\psi^{Spatial}_{[1,1,0,0,0,0]}$, $\psi^{Spatial}_{[1,0,1,0,0,0]}$, $\psi^{Spatial}_{[0,1,1,0,0,0]}$.
	\end{item}
	\\
%%%%%%%%%%%%%%%%%%%%%%%%%%%%%%%%%%%%%%%%%%
	\begin{item}4th Quanta ($Q=6$)\\
	$\psi^{Spatial}_{[3,0,0,0,0,0]}$, $\psi^{Spatial}_{[0,3,0,0,0,0]}$, $\psi^{Spatial}_{[0,0,3,0,0,0]}$,
	\\
	$\psi^{Spatial}_{[2,0,0,1,1,0]}$, $\psi^{Spatial}_{[0,2,0,1,1,0]}$, $\psi^{Spatial}_{[0,0,2,1,1,0]}$,
	\\
	$\psi^{Spatial}_{[2,0,0,1,0,1]}$, $\psi^{Spatial}_{[0,2,0,1,0,1]}$, $\psi^{Spatial}_{[0,0,2,1,0,1]}$,
	\\
	$\psi^{Spatial}_{[2,0,0,0,1,1]}$, $\psi^{Spatial}_{[0,2,0,0,1,1]}$, $\psi^{Spatial}_{[0,0,2,0,1,1]}$,
	\\
	$\psi^{Spatial}_{[1,1,0,1,1,0]}$, $\psi^{Spatial}_{[1,0,1,1,1,0]}$, $\psi^{Spatial}_{[0,1,1,1,1,0]}$,
	\\
	$\psi^{Spatial}_{[1,1,0,1,0,1]}$, $\psi^{Spatial}_{[1,0,1,1,0,1]}$, $\psi^{Spatial}_{[0,1,1,1,0,1]}$,
	\\
	$\psi^{Spatial}_{[1,1,0,0,1,1]}$, $\psi^{Spatial}_{[1,0,1,0,1,1]}$, $\psi^{Spatial}_{[0,1,1,0,1,1]}$,
	\\
	$\psi^{Spatial}_{[1,0,0,2,2,0]}$, $\psi^{Spatial}_{[0,1,0,2,2,0]}$, $\psi^{Spatial}_{[0,0,1,2,2,0]}$,
	\\
	$\psi^{Spatial}_{[1,0,0,2,0,2]}$, $\psi^{Spatial}_{[0,1,0,2,0,2]}$, $\psi^{Spatial}_{[0,0,1,2,0,2]}$,
	\\
	$\psi^{Spatial}_{[1,0,0,0,2,2]}$, $\psi^{Spatial}_{[0,1,0,0,2,2]}$, $\psi^{Spatial}_{[0,0,1,0,2,2]}$,
	\\
	$\psi^{Spatial}_{[0,0,0,3,3,0]}$, $\psi^{Spatial}_{[0,0,0,3,0,3]}$, $\psi^{Spatial}_{[0,0,0,0,3,3]}$,
	\\
	$\psi^{Spatial}_{[1,0,0,1,1,2]}$, $\psi^{Spatial}_{[0,1,0,1,1,2]}$, $\psi^{Spatial}_{[0,0,1,1,1,2]}$,
	\\
	$\psi^{Spatial}_{[1,0,0,1,2,1]}$, $\psi^{Spatial}_{[0,1,0,1,2,1]}$, $\psi^{Spatial}_{[0,0,1,1,2,1]}$,
	\\
	$\psi^{Spatial}_{[1,0,0,2,1,1]}$, $\psi^{Spatial}_{[0,1,0,2,1,1]}$, $\psi^{Spatial}_{[0,0,1,2,1,1]}$,
	\\
	$\psi^{Spatial}_{[1,2,0,0,0,0]}$, $\psi^{Spatial}_{[2,1,0,0,0,0]}$,
	\\
	$\psi^{Spatial}_{[1,0,2,0,0,0]}$, $\psi^{Spatial}_{[2,0,1,0,0,0]}$,
	\\
	$\psi^{Spatial}_{[0,1,2,0,0,0]}$, $\psi^{Spatial}_{[0,2,1,0,0,0]}$,
	\\
	$\psi^{Spatial}_{[1,1,1,0,0,0]}$.
	\end{item}
	\\
%%%%%%%%%%%%%%%%%%%%%%%%%%%%%%%%%%%%%%%%%%
	\begin{item}5th Quanta ($Q=8$)\\
	$\psi^{Spatial}_{[4,0,0,0,0,0]}$, $\psi^{Spatial}_{[0,4,0,0,0,0]}$, $\psi^{Spatial}_{[0,0,4,0,0,0]}$,
	\\
	$\psi^{Spatial}_{[3,0,0,1,1,0]}$, $\psi^{Spatial}_{[0,3,0,1,1,0]}$, $\psi^{Spatial}_{[0,0,3,1,1,0]}$,
	\\$\psi^{Spatial}_{[3,0,0,1,0,1]}$, $\psi^{Spatial}_{[0,3,0,1,0,1]}$, $\psi^{Spatial}_{[0,0,3,1,0,1]}$,
	\\$\psi^{Spatial}_{[3,0,0,0,1,1]}$, $\psi^{Spatial}_{[0,3,0,0,1,1]}$, $\psi^{Spatial}_{[0,0,3,0,1,1]}$,
	\\
	$\psi^{Spatial}_{[1,2,0,1,1,0]}$, $\psi^{Spatial}_{[2,1,0,1,1,0]}$,
	\\
	$\psi^{Spatial}_{[1,0,2,1,1,0]}$, $\psi^{Spatial}_{[2,0,1,1,1,0]}$,
	\\
	$\psi^{Spatial}_{[0,1,2,1,1,0]}$, $\psi^{Spatial}_{[0,2,1,1,1,0]}$,
	\\
	$\psi^{Spatial}_{[1,2,0,1,0,1]}$, $\psi^{Spatial}_{[2,1,0,1,0,1]}$,
	\\
	$\psi^{Spatial}_{[1,0,2,1,0,1]}$, $\psi^{Spatial}_{[2,0,1,1,0,1]}$,
	\\
	$\psi^{Spatial}_{[0,1,2,1,0,1]}$, $\psi^{Spatial}_{[0,2,1,1,0,1]}$,
	\\
	$\psi^{Spatial}_{[1,2,0,0,1,1]}$, $\psi^{Spatial}_{[2,1,0,0,1,1]}$,
	\\
	$\psi^{Spatial}_{[1,0,2,0,1,1]}$, $\psi^{Spatial}_{[2,0,1,0,1,1]}$,
	\\
	$\psi^{Spatial}_{[0,1,2,0,1,1]}$, $\psi^{Spatial}_{[0,2,1,0,1,1]}$,
	\\
	$\psi^{Spatial}_{[1,1,1,1,1,0]}$, $\psi^{Spatial}_{[1,1,1,1,0,1]}$, $\psi^{Spatial}_{[1,1,1,0,1,1]}$,
	\\
	$\psi^{Spatial}_{[2,0,0,2,2,0]}$, $\psi^{Spatial}_{[0,2,0,2,2,0]}$, $\psi^{Spatial}_{[0,0,2,2,2,0]}$,
	\\$\psi^{Spatial}_{[2,0,0,2,0,2]}$, $\psi^{Spatial}_{[0,2,0,2,0,2]}$, $\psi^{Spatial}_{[0,0,2,2,0,2]}$,
	\\$\psi^{Spatial}_{[2,0,0,0,2,2]}$, $\psi^{Spatial}_{[0,2,0,0,2,2]}$, $\psi^{Spatial}_{[0,0,2,0,2,2]}$,	
	\\
	$\psi^{Spatial}_{[1,1,0,2,2,0]}$, $\psi^{Spatial}_{[1,0,1,2,2,0]}$, $\psi^{Spatial}_{[1,0,1,2,2,0]}$,
	\\$\psi^{Spatial}_{[1,1,0,2,0,2]}$, $\psi^{Spatial}_{[1,0,1,2,0,2]}$, $\psi^{Spatial}_{[1,0,1,2,0,2]}$,
	\\$\psi^{Spatial}_{[1,1,0,0,2,2]}$, $\psi^{Spatial}_{[1,0,1,0,2,2]}$, $\psi^{Spatial}_{[0,1,1,0,2,2]}$,
	\\
	$\psi^{Spatial}_{[1,0,0,3,3,0]}$, $\psi^{Spatial}_{[0,1,0,3,3,0]}$, $\psi^{Spatial}_{[0,0,1,3,3,0]}$,
	\\
	$\psi^{Spatial}_{[1,0,0,3,0,3]}$, $\psi^{Spatial}_{[0,1,0,3,0,3]}$, $\psi^{Spatial}_{[0,0,1,3,0,3]}$,
	\\
	$\psi^{Spatial}_{[1,0,0,0,3,3]}$, $\psi^{Spatial}_{[0,1,0,0,3,3]}$, $\psi^{Spatial}_{[0,0,1,0,3,3]}$,
	\\
	$\psi^{Spatial}_{[2,0,0,1,1,2]}$, $\psi^{Spatial}_{[0,2,0,1,1,2]}$, $\psi^{Spatial}_{[0,0,2,1,1,2]}$,
	\\
	$\psi^{Spatial}_{[2,0,0,1,2,1]}$, $\psi^{Spatial}_{[0,2,0,1,2,1]}$, $\psi^{Spatial}_{[0,0,2,1,2,1]}$,
	\\
	$\psi^{Spatial}_{[2,0,0,2,1,1]}$, $\psi^{Spatial}_{[0,2,0,2,1,1]}$, $\psi^{Spatial}_{[0,0,2,2,1,1]}$,
	\\
	$\psi^{Spatial}_{[1,1,0,1,1,2]}$, $\psi^{Spatial}_{[1,0,1,1,1,2]}$, $\psi^{Spatial}_{[0,1,1,1,1,2]}$,
	\\
	$\psi^{Spatial}_{[1,1,0,1,2,1]}$, $\psi^{Spatial}_{[1,0,1,1,2,1]}$, $\psi^{Spatial}_{[0,1,1,1,2,1]}$,
	\\
	$\psi^{Spatial}_{[1,1,0,2,1,1]}$, $\psi^{Spatial}_{[1,0,1,2,1,1]}$, $\psi^{Spatial}_{[0,1,1,2,1,1]}$,
	\\
	$\psi^{Spatial}_{[2,2,0,0,0,0]}$, $\psi^{Spatial}_{[2,0,2,0,0,0]}$, $\psi^{Spatial}_{[0,2,2,0,0,0]}$,
	\\
	$\psi^{Spatial}_{[1,3,0,0,0,0]}$, $\psi^{Spatial}_{[3,1,0,0,0,0]}$,
	\\
	$\psi^{Spatial}_{[1,0,3,0,0,0]}$, $\psi^{Spatial}_{[3,0,1,0,0,0]}$,
	\\
	$\psi^{Spatial}_{[0,1,3,0,0,0]}$, $\psi^{Spatial}_{[0,3,1,0,0,0]}$,
	\\
	$\psi^{Spatial}_{[1,1,2,0,0,0]}$, $\psi^{Spatial}_{[1,2,1,0,0,0]}$, $\psi^{Spatial}_{[2,1,1,0,0,0]}$.
	\end{item}
\end{itemize}
{}\
\\
\\
As can be seen in the above classification, we have considered bases with internal angular momenta up to the $l_i=3$ states because the contribution from the higher angular momentum basis is not so significant. There are other types of harmonic oscillator bases constituting the complete set, such as $\psi^{Spatial}_{[0,0,0,1,1,1]}$.  However, there is no transitional matrix element of the Hamiltonian between this type of bases and the other bases presented above. Thus, they do not contribute to the ground state masses of the tetraquarks. In doing so, we have observed the convergence behavior of the ground state mass, and have included the harmonic oscillator bases up to the 5th quanta in the calculations.

On the other hand, in order to satisfy the Pauli principle, it is necessary to consider the permutation symmetries of the spatial functions as well as the other parts of the wave function. In the Jacobi coordinates set 1 of our reference, it is obvious that $T_{QQ}$ configuration has symmetries under the permutations (12) and (34). $T_{QQ'}$ has symmetry under the permutation (12), while $us\bar{b}\bar{b}$ has symmetry under the permutation (34) as summarized in Table~\ref{SymmetryCSF}. The symmetry property of the spatial functions can be summarized for each tetraquark configuration in Table \ref{SymmetrySpatial}.

%%%%%%%%%%%%%%%%%%%%%		Table 4. Symmetries of Spatial Function		%%%%%%%%%%%%%%%%%%%%%%%%%
\begin{widetext}

\begin{table}[t]

\caption{The permutation symmetry properties of the spatial functions and the corresponding CS bases for each tetraquark configuration. In the table, $(12)$ and $(34)$ indicate the permutations. $T_{QQ'}$ stands for $ud\bar{Q}\bar{Q'}$. Here, $+1(-1)$ indicates that the spatial function is symmetric (antisymmetric) under the corresponding permutations. The blank in the table implies that there is no symmetry constraint under the corresponding permutation.}	
\centering
\begin{tabular}{ccccccccccccccc}
\hline
\hline	\multirow{2}{*}{Type}	&&	\multicolumn{3}{c}{$T_{QQ}$}	&&&	\multicolumn{3}{c}{$T_{QQ'}$}	&&&	\multicolumn{3}{c}{$us\bar{b}\bar{b}$}	\\
\cline{3-5}\cline{8-10}\cline{13-15}
							&&	(12)	&	(34)	&	CS bases	&&&	(12)	&	(34)	&	CS bases	&&&	(12)	&	(34)	&	CS bases	\\
\hline 
$l_1=l_2=$ odd, $l_3=$ even	&&	$-1$	&	$-1$	&	$\psi^{CS}_4$, $\psi^{CS}_5$	&&&	$-1$	&		&	$\psi^{CS}_2$, $\psi^{CS}_4$, $\psi^{CS}_5$	&&&			&	$-1$&	$\psi^{CS}_1$, $\psi^{CS}_4$, $\psi^{CS}_5$		\\
$l_1=l_3=$ odd, $l_2=$ even	&&	$-1$	&	+1		&		$\psi^{CS}_2$			&&&	$-1$	&		&	$\psi^{CS}_2$, $\psi^{CS}_4$, $\psi^{CS}_5$	&&&			&	+1	&	$\psi^{CS}_2$, $\psi^{CS}_3$, $\psi^{CS}_6$		\\
$l_2=l_3=$ odd, $l_1=$ even	&&	+1		&	$-1$	&		$\psi^{CS}_1$			&&&	+1		&		&	$\psi^{CS}_1$, $\psi^{CS}_3$, $\psi^{CS}_6$	&&&			&	$-1$&	$\psi^{CS}_1$, $\psi^{CS}_4$, $\psi^{CS}_5$		\\
\\
$l_1=l_2=l_3=$ even			&&	+1		&	+1		&	$\psi^{CS}_3$, $\psi^{CS}_6$	&&&	+1		&		&	$\psi^{CS}_1$, $\psi^{CS}_3$, $\psi^{CS}_6$	&&&			&	+1	&	$\psi^{CS}_2$, $\psi^{CS}_3$, $\psi^{CS}_6$		\\
\hline 
\hline
\label{SymmetrySpatial}
\end{tabular}
\end{table}
%%%%%%%%%%%%%%%%%%%%%%%%%		Table 5. Tetraquark Results		%%%%%%%%%%%%%%%%%%%%%%%%%%%%%
\begin{table}[t]
\caption{The masses and binding energies $B_{T}$ of the tetraquark states obtained with the  fitting parameters in Eq.~(\ref{FitParameters}). The binding energy $B_{T}$ is defined by the difference between the tetraquark mass and the sum of the masses of the lowest threshold mesons, $B_{T} \equiv M_{Tetraquark} - M_{meson 1} - M_{meson 2}$. The values in the parentheses are the results in our previous work\cite{Woosung:NPA2019}.}
\begin{tabular}{ccccccccc}
\hline
\hline
Type					&&	($I$, $S$)	&&	Mass(MeV)	&&	Variational Parameters(fm$^{-2}$)	&&	$B_{T}$(MeV)	\\
\hline 
%%% Tbb %%%%%
$ud\bar{b}\bar{b}$	&&	(0,1)	&&	10517 (10518)	&&	$a_{1}= 3.9 (2.8), \, a_{2}= 25.0 (20.9), \, a_{3}= 3.8 (2.8)$	&&	$-145$ ($-121$)	\\
%%% Tcc %%%%%
$ud\bar{c}\bar{c}$	&&	(0,1)	&&	3873 (3965)	&&	$a_{1}= 2.6 (2.8), \, a_{2}= 4.6 (7.6), \, a_{3}= 4.6 (2.7)$	&&	+13 (+99)	\\
%%% Tcb %%%%%
$ud\bar{c}\bar{b}$	&&	(0,1)	&&	7212 (7262)	&&	$a_{1}= 3.1 (3.1), \, a_{2}= 8.0 (10.3), \, a_{3}= 5.0 (2.7)$	&&	$-3$ (+49)	\\
%%% usbb %%%%
$us\bar{b}\bar{b}$	&&	(1/2,1)	&&	10694 (10684)	&&	$a_{1}= 4.0 (3.5), \, a_{2}= 21.4 (20.6), \, a_{3}= 6.0 (3.5)$	&&	$-42$ ($-7$)	\\
%%% Tsc %%%%
$ud\bar{s}\bar{c}$	&&	(0,1)	&&		2596	&&	$a_{1}= 2.4, \, a_{2}= 3.9, \, a_{3}= 5.3$	&&	+91	\\
%%% Tsc %%%%
$ud\bar{s}\bar{c}$	&&	(0,2)	&&	2938	&&	$a_{1}= 1.6, \, a_{2}= 2.1, \, a_{3}= 4.0$	&&	+57	\\
%%% Tsb %%%%
$ud\bar{s}\bar{b}$	&&	(0,1)	&&	5949	&&	$a_{1}= 2.6, \, a_{2}= 5.1, \, a_{3}= 6.3$	&&	+90	\\
%%% Tsb %%%%
$ud\bar{s}\bar{b}$	&&	(0,2)	&&	6298	&&	$a_{1}= 1.8, \, a_{2}= 2.6, \, a_{3}= 4.9$	&&	+63	\\
\hline 
\hline
\label{result}
\end{tabular}
\end{table}

\end{widetext}
%%%%%%%%%%%%%%%%%%%%%%%%%%%%%%%%%%%%%%%%%%%%%%%%%%%%%%%%%%%%%%%%%%%%%%%%%%%%%%%%%%%%%%%%%%%%%
Note that there is antisymmetric part in some of the spatial bases due to the symmetry property of the harmonic oscillator bases as seen in Table.~\ref{SymmetrySpatial}. Thus, contrary to the previous work\cite{Woosung:NPA2019} where only the fully symmetric spatial basis $\psi^{Spatial}_{[0,0,0,0,0,0]}$ was used in the calculations, all the CS bases are used in the present work.
%%%%%%%%%%%%%%%%%%%%%%%%%%%%%%%%%%%%%%%%%%%%%%%%%%%%%%%%%%%%%%%%%%%%%%%%%%%%%%%%%%%%%%%%%%%%%%%
\section{Numerical Analysis}
%%%%%%%%%%%%%%%%%%%%%%%%%%%%%%%%%%%%%%%%%%%%%%%%%%%%%%%%%%%%%%%%%%%%%%%%%%%%%%%%%%%%%%%%%%%%%%%
We use the total wave function discussed in the previous section, and perform a variational method to determine the ground state masses for the tetraquarks. The results are shown in Table~\ref{result}. We found that $ud\bar{b}\bar{b}$ and $us\bar{b}\bar{b}$ are stable against the lowest strong decay threshold. In particular, one can see that the effect from the harmonic oscillator bases appears to lower the binding energies. As a result, $T_{cb} (ud\bar{c}\bar{b})$ is found to be below the lowest threshold. One of the major contributions to this is coming from the excited orbital states.
 
In this section, we first discuss a similar tendency of the ground state wave functions between the tetraquark and the meson structures. Then we will discuss the effects of the excited orbital states.  
Also, we investigate the spatial distibutions and the sizes for the tetraquarks. Finally, we compare our model with other works.

%%%%%%%%%%%%%%%%%%%%%%%%%%%%%%%%%%%%%%%%%%%%%%%%%%%%%%%%%%%%%%%%%%%%%%%%%%%%%%%%%%%%%%%%%%%%%%%
\subsection{Ground State of Tetraquarks}
%%%%%%%%%%%%%%%%%%%%%%%%%%%%%%%%%%%%%%%%%%%%%%%%%%%%%%%%%%%%%%%%%%%%%%%%%%%%%%%%%%%%%%%%%%%%%%%
We first analyze the ground state of the tetraquarks by expanding the wave function in terms of the complete set of harmonic oscillator bases.  Using the variational method, we find that the expansion coefficients of the bases rapidly converge to zero.

\begin{figure}[h!]
	\centering
	
	\includegraphics[width=0.9\columnwidth]{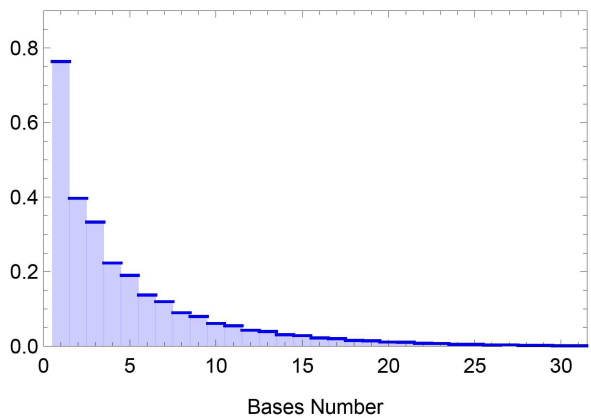}
    \caption{The expansion coefficients of the ground state wave function for $D$ meson. The total number of the Hamiltonian bases (the total wave functions) is 31, each of which is categorized according to a given quanta in the meson structure.}
    \label{DGround}
\end{figure}

\begin{figure}[h!]
	\centering
	
	\includegraphics[width=0.95\columnwidth]{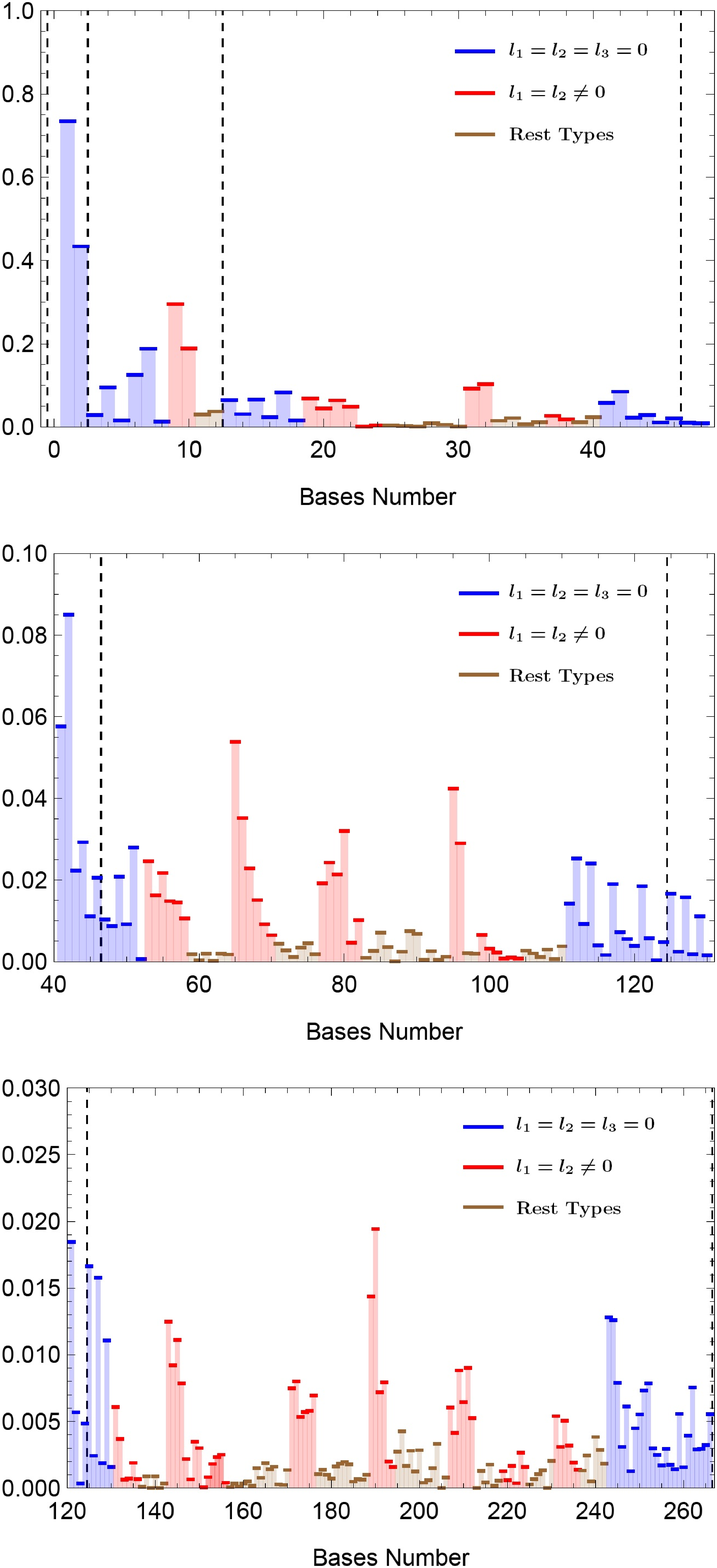}
    \caption{Same as Figure~\ref{DGround} but for $T_{cc}$. The figure is separated into three parts to clearly see the coefficients with appropriate scales. The dashed-lines separate them into each quanta of the bases. The total number of Hamiltonian bases is 266 for $T_{cc}$, which are composed of the five quanta of harmonic oscillator bases as listed in Section \ref{Section3C}.}
    \label{TccGround}
\end{figure}

\begin{figure}[h!]
	\centering
	
	\includegraphics[width=0.95\columnwidth]{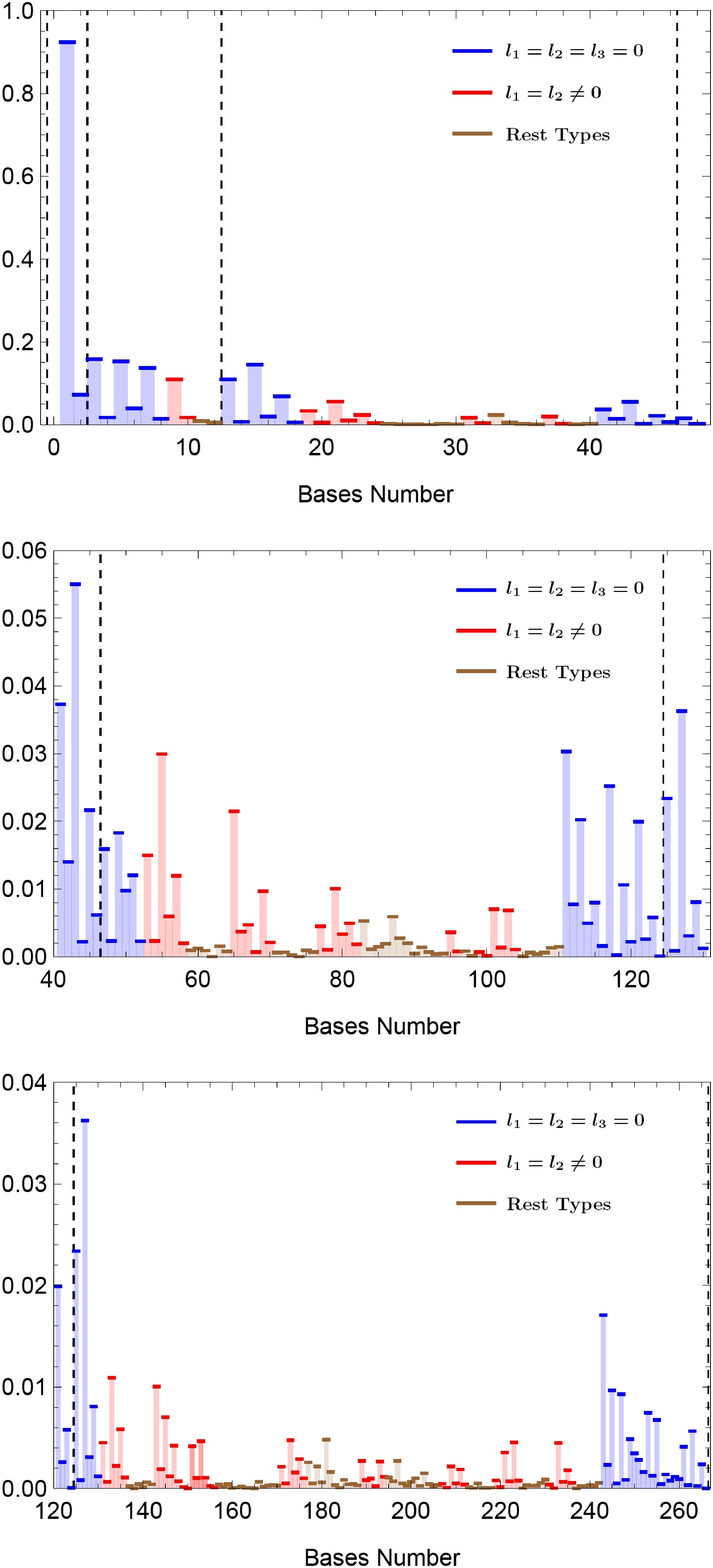}
	\caption{Same as Figure~\ref{TccGround} but for $T_{bb}$. The Hamiltonian bases for $T_{bb}$ are the same with those for $T_{cc}$.}
    \label{TbbGround}
\end{figure}

\begin{figure}[t]
	\centering
	
	\includegraphics[width=0.95\columnwidth]{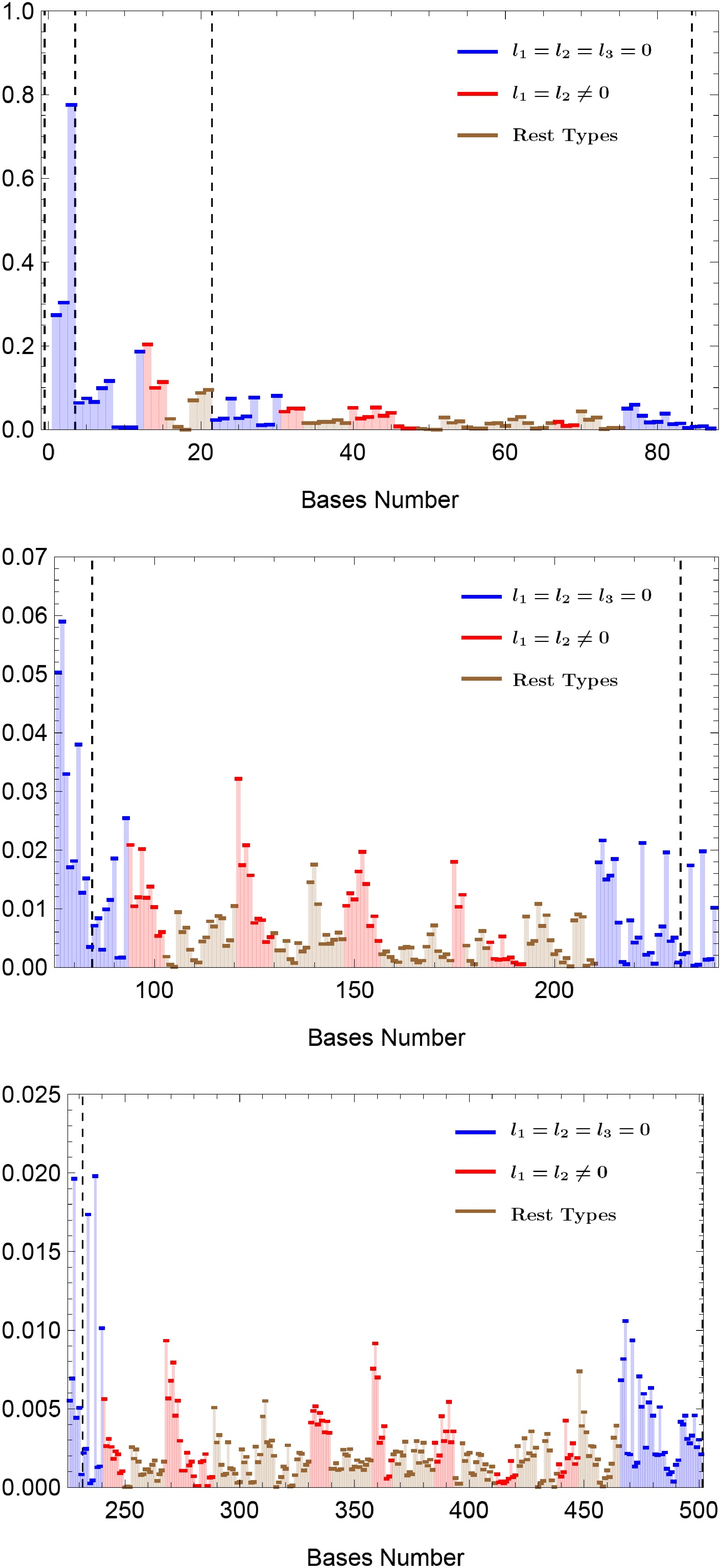}
	\caption{Same as Figure~\ref{TccGround} but for $T_{cb}$. The number of Hamiltonian bases for $T_{cb}$ is 501, but the spatial bases are the same with those for $T_{cc}$.}
    \label{TcbGround}
\end{figure}

\begin{figure}[t]
	\centering
	
	\includegraphics[width=0.95\columnwidth]{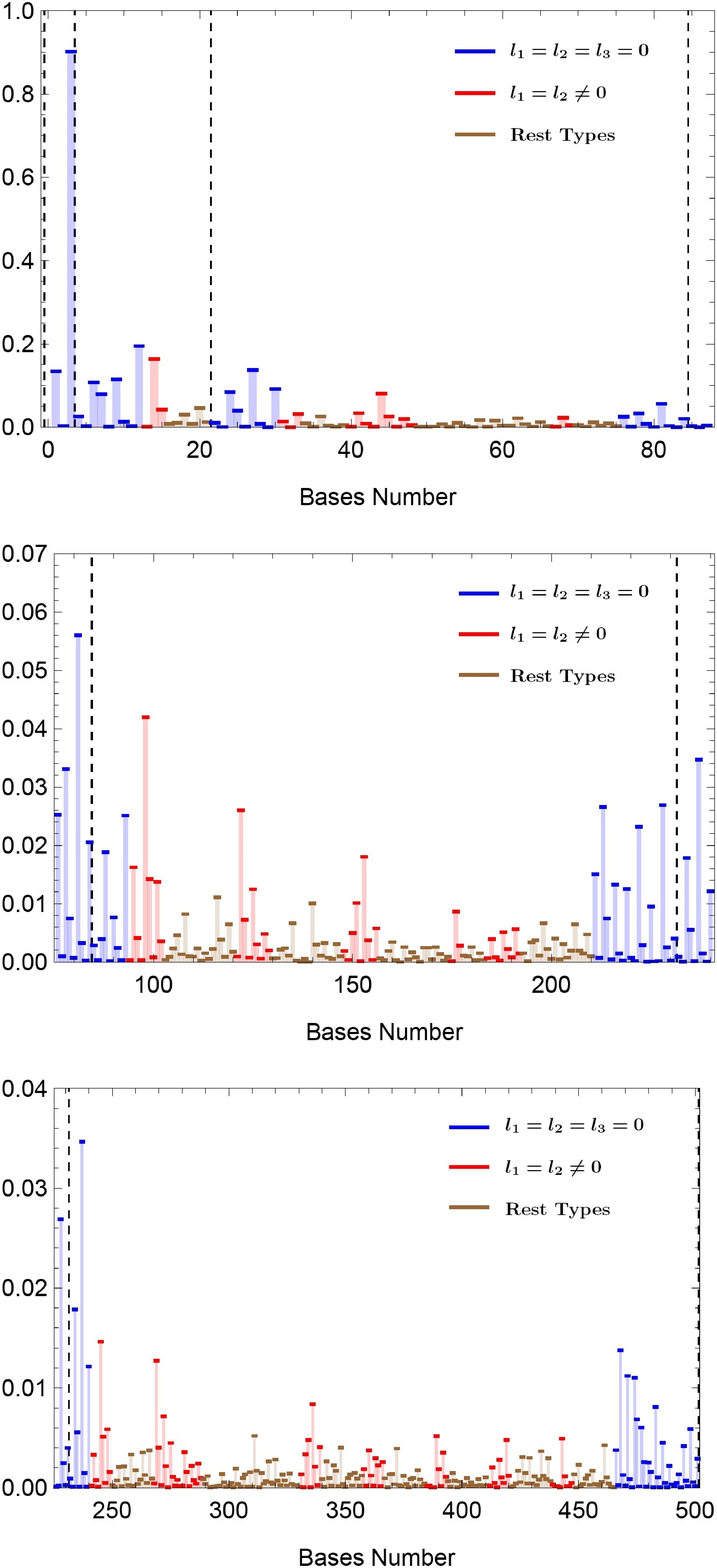}
	\caption{Same as Figure~\ref{TccGround} but for $us\bar{b}\bar{b}$. The number of Hamiltonian bases for $us\bar{b}\bar{b}$ is 501, but the spatial bases are the same with those for $T_{cc}$.}
    \label{usbbGround}
\end{figure}

As discussed in Section~\ref{Section3C}, each of the harmonic oscillator bases for mesons is classified by different quanta. Then, as can be seen in Figure~\ref{DGround} for the $D$ meson, the expansion coefficient monotonically decreases as desired when the quanta of the bases increases. This tendency can be seen also for the tetraquark states in Figures~\ref{TccGround}-\ref{usbbGround}, respectively. In addition to the convergence of the expansion coefficients, one can observe the convergence of the ground state masses as the number of the harmonic oscillator bases and their quanta increase as seen in Figure~\ref{Dmeson} for the $D$ meson. We then determined the ground state masses for the mesons, baryons, and the tetraquarks when the convergent values change by just few MeV for

%%%%%%%%%%  Table 6. Contributions from the angular bases for Tcc and Tcb %%%%%%%%%%%%%%%%%%%%
\begin{widetext}

\begin{table}[t]

\caption{The changes in masses of the tetraquarks when the indicated spatial bases are included in the calculations with the corresponding dominant CS bases.}	
\centering
\begin{tabular}{ccccccccccccc}
\hline
\hline	Spatial Bases						&&&		$M_{T_{bb}}$ (MeV)	&&&	$M_{T_{cc}}$ (MeV)	&&&	$M_{T_{cb}}$ (MeV)	&&&	$M_{us\bar{b}\bar{b}}$ (MeV)	\\
\hline
$\psi^{Spatial}_{[0,0,0,0,0,0]}$										
&&&			10578		&&&			4002		&&&		7316		&&&	10764	\\
\\
$\psi^{Spatial}_{[0,0,0,0,0,0]}$, $\psi^{Spatial}_{[0,0,0,1,1,0]}$	
&&&			10567		&&&			3968		&&&		7294		&&&	10748	\\
\\
$\psi^{Spatial}_{[0,0,0,0,0,0]}$, $\psi^{Spatial}_{[0,0,0,1,1,0]}$	
&&&	\multirow{2}{*}{10566}&&&	\multirow{2}{*}{3967}&&&	\multirow{2}{*}{7290}&&&	\multirow{2}{*}{10746}	\\
$\psi^{Spatial}_{[0,0,0,1,0,1]}$, $\psi^{Spatial}_{[0,0,0,0,1,1]}$
\\
\hline
	Total Change in Mass		
&&&			-12			&&&			-35			&&&		-26		&&&		-18	\\
\hline
\hline 
\label{AngularContribution}
\end{tabular}
\end{table}

\end{widetext}
%%%%%%%%%%%%%%%%%%%%%%%%%%%%%%%%%%%%%%%%%%%%%%%%%%%%%%%%%%%%%%%%%%%%%%%%%%%%%%%%%%%%%%%%%%%%%%%

%%%%%%%%%%%%%%%%%%%%%		Figures of Tetraquark Ground Wave Functions	%%%%%%%%%%%%%%%%%%%%%
%%%%%%%%%%%%%%%%%%%%%%%%%%%%%%%%%%%%%%%%%%%%%%%%%%%%%%%%%%%%%%%%%%%%%%%%%%%%%%%%%%%%%%%%%%%%%%%
{}\
\\
the entire bases contained in the last quanta, which turns out to be the 5th quanta. On the other hand, the convergence behavior of the expansion coefficients for the tetraquarks is not monotonic. This is so because while the nonzero orbital bases are ordered later, their contributions to the tetraquark configurations are important due to the attraction coming from the dipole and quadrupole moments.  Still the average value of the coefficients in each quanta monotonically decreases.

Discussing the tetraquark state in some more detail, one finds that from the second quanta on, many of the largest coefficients correspond to the harmonic oscillator bases with $l_1=l_2\neq0$, which implies that the contributions from this type of bases are important to obtain the exact ground state masses for the tetraquarks.
This can be quantitatively seen by the mass changes when this type of bases are additively included in the calculations. For simplicity, considering only the dominant CS bases, we first evaluate the mass with only the spatial basis $\psi^{Spatial}_{[0,0,0,0,0,0]}$, and then compare this with the mass obtained by adding the contribution from the basis $\psi^{Spatial}_{[0,0,0,1,1,0]}$. The results for such a calculation for some of the tetraquarks are summarized in Table~\ref{AngularContribution}.
The contribution from the basis $\psi^{Spatial}_{[0,0,0,1,1,0]}$ appears to lower the tetraquark masses by 11 MeV in $T_{bb}$ and 34 MeV in $T_{cc}$, respectively. The effect is larger in $T_{cc}$ because the magnitudes of the corresponding expansion coefficients in $T_{cc}$ are obviously larger than the others in the second quanta, while it is not so in $T_{bb}$. 

Also, considering only the dominant CS bases, the coefficient in the first quanta corresponding to the spatial basis $\psi^{Spatial}_{[0,0,0,0,0,0]}$ is larger in $T_{bb}$ than that in $T_{cc}$. The first coefficient in $T_{bb}$ is 0.94, which implies that there are less contributions from the other bases in the tetraquarks of heavier antiquarks than in the tetraquarks of lighter antiquarks. However, compared to the total changes in masses from the value only with the first basis in Table~\ref{AngularContribution} to the exact ground state masses in Table~\ref{result}, the contribution from the basis $\psi^{Spatial}_{[0,0,0,1,1,0]}$ is still important in $T_{bb}$ as well.

Let us now discuss the coefficients corresponding to the rest types of excited orbital bases. For $T_{cc}$ in Figure~\ref{TccGround} and $T_{cb}$ in Figure~\ref{TcbGround}, the coefficients in $T_{cb}$ are not so small relative to those in $T_{cc}$. Obviously in the second quanta, it is comparable to the others in $T_{cb}$, while they are not so in $T_{cc}$. As a result, as can be seen in Table~\ref{AngularContribution}, the mass is lowered by 1 MeV in $T_{cc}$ but it is 4 MeV in $T_{cb}$ when adding the bases $\psi^{Spatial}_{[0,0,0,1,0,1]}$ and $\psi^{Spatial}_{[0,0,0,0,1,1]}$ to the calculations. This is due to the symmetry breaking in the flavor part of antiquarks, relative to $T_{cc}$ structure. A similar behavior appears also in the comparison between $T_{bb}$ and $us\bar{b}\bar{b}$ depicted in Figures~\ref{TbbGround}, \ref{usbbGround}. In this case, the flavor symmetry is bronken in the quark part. Comparing the changes in masses between $T_{bb}$ and $us\bar{b}\bar{b}$, the contributions from the bases $\psi^{Spatial}_{[0,0,0,1,0,1]}$ and $\psi^{Spatial}_{[0,0,0,0,1,1]}$ are little larger in $us\bar{b}\bar{b}$. Therefore, all the types of the excited orbital states are necessary to obtain the exact ground state masses for the tetraquark states.
 
%%%%%%%%%%%%%%%%%%%%%%%%%%%%%%%%%%%%%%%%%%%%%%%%%%%%%%%%%%%%%%%%%%%%%%%%%%%%%%%%%%%%%%%%%%%%%%%
\subsection{Spatial Size of Tetraquarks}
\label{SizeSec}
%%%%%%%%%%%%%%%%%%%%%%%%%%%%%%%%%%%%%%%%%%%%%%%%%%%%%%%%%%%%%%%%%%%%%%%%%%%%%%%%%%%%%%%%%%%%%%%
It is also useful to investigate the spatial size of the tetraquarks, and the spatial distribution of the constituent quarks in the tetraquark structure. The relative distances between the quarks are listed in Table~\ref{RelDistances}.
%%%%%%%%%%%%%%%%%%%%%%%%		Table 7. Relative Distances		%%%%%%%%%%%%%%%%%%%%%%%%%%%%%%%%%%
\begin{table}[h!]

\caption{The relative distances between the quarks in the tetraquarks in fm unit. The distances are obtained with the ground state of the tetraquakrs. In the table, $(i,j)$ denotes the distance between $i$ and $j$ quarks, and $(1,2)$-$(3,4)$ that between the center of masses of the pairs $(1,2)$ and $(3,4)$.}	
\centering
\begin{tabular}{ccccccccccc}
\hline
\hline
	Quark Pair		&&&&	$T_{bb}$	&&	$T_{cc}$	&&	$T_{cb}$	&&	$us\bar{b}\bar{b}$	\\
\hline
		$(1,2)$		&&&&	0.676	&&	0.830	&&	0.753	&&	0.644	\\
		$(1,3)$		&&&&	0.592	&&	0.672	&&	0.631	&&	0.584	\\
		$(1,4)$		&&&&	0.592	&&	0.672	&&	0.612	&&	0.584	\\
		$(2,3)$		&&&&	0.592	&&	0.672	&&	0.631	&&	0.490	\\
		$(2,4)$		&&&&	0.592	&&	0.672	&&	0.612	&&	0.490	\\
		$(3,4)$		&&&&	0.268	&&	0.610	&&	0.464	&&	0.287	\\
\\
	$(1,2)$-$(3,4)$		&&&&	0.463	&&	0.433	&&	0.441	&&	0.397	\\
\hline 
\hline
\label{RelDistances}
\end{tabular}
\end{table}
%%%%%%%%%%%%%%%%%%%%%%%%		Table 8. Threshold Meson Size	%%%%%%%%%%%%%%%%%%%%%%%%%%%%%%%%%
\begin{table}[h!]

\caption{The spatial sizes of the lowest decay threshold mesons for each tetraquark state in fm unit. In the last columns for each section, ``Total'' indicates the sum of the sizes of two threshold mesons.}	
\centering

\begin{tabular}{cccccccccccc}
\hline
\hline
	Tetraquarks	&&	\multicolumn{4}{c}{$T_{bb}$}	&&&	\multicolumn{4}{c}{$T_{cc}$}			\\
\cline{3-6}\cline{9-12}
Lowest Threshold	&&	$B$		&	$B^*$	&&	Total	&&&		$D$		&		$D^*$	&&	Total	\\
\hline
	Size			&&	0.525	&	0.551	&&	1.076	&&&		0.519	&		0.586	&&	1.105	\\
\hline
\hline
	Tetraquarks	&&	\multicolumn{4}{c}{$T_{cb}$}	&&&	\multicolumn{4}{c}{$us\bar{b}\bar{b}$}		\\
\cline{3-6}\cline{9-12}
Lowest Threshold	&&	$D$		&	$B^*$	&&	Total	&&&		$B_s$	&		$B^*$	&&	Total	\\
\hline
	Size			&&	0.519	&	0.551	&&	1.070	&&&		0.397	&		0.551	&&	0.948	\\
\hline 
\hline
\label{MesonSize}
\end{tabular}
\end{table}
%%%%%%%%%%%%%%%%%%%%%%%%%%%%%%%%%%%%%%%%%%%%%%%%%%%%%%%%%%%%%%%%%%%%%%%%%%%%%%%%%%%%%%%%%%%%%%%
The relative distance between the heavier quarks are in general closer than that of the lighter quarks\cite{Woosung:NPA2019}. This tendency is maintained in each tetraquark state as seen in Table~\ref{RelDistances}. Looking into the relative distances, it is found that the relative distances except for $(1,2)$ and $(3,4)$ pairs are all the same in $T_{QQ}$ structure. For $T_{QQ'}$ structure, the relative distances for the pairs $(1,3)$ and $(2,3)$ are the same, and also for the pairs $(1,4)$ and $(2,4)$. Likewise, the relative distances for the pairs $(1,3)$ and $(1,4)$ are the same as are those for the pairs $(2,3)$ and $(2,4)$ in $us\bar{b}\bar{b}$. This is  due to the flavor symmetry in each tetraquark structure, and can be simply evaluated through the permutation symmetry for the ground state wave function in each tetraquark state.
Since the total wave function satisfies the Pauli principle in each tetraquark, if we denote the ground state wave function by $|\Psi^{Tetraquark}_G \rangle \equiv |\psi^{Spatial}\rangle \times |\psi^{CS}\rangle$, the permutation symmetries for each ground state are as follows.
\begin{small}
\begin{eqnarray}
&(12) \left| \Psi^{T_{QQ}}_G \right\rangle = (34) \left|\Psi^{T_{QQ}}_G \right\rangle = - \left| \Psi^{T_{QQ}}_G \right\rangle & \quad \, {\rm for} \,\,\, T_{QQ}\,,
\nonumber \\
&(12) \left|\Psi^{T_{QQ'}}_G \right\rangle = - \left| \Psi^{T_{QQ'}}_G \right\rangle & \, \quad {\rm for} \,\,\, T_{QQ'} \,,
\nonumber \\
&(34) \left|\Psi^{us\bar{b}\bar{b}}_G \right\rangle = - \left| \Psi^{us\bar{b}\bar{b}}_G \right\rangle & \quad \, {\rm for} \,\,\, us\bar{b}\bar{b} \,.
\nonumber
\end{eqnarray}
\end{small}
\!\!Then the relative distances in each tetraquark structure can be obtained as follows.
\\
For $T_{QQ}$,
\begin{eqnarray}
&&\left\langle \Psi^{T_{QQ}}_G \right| (12)^{-1}(12) |\mathbf{r}_1 - \mathbf{r}_3| (12)^{-1}(12) \left| \Psi^{T_{QQ}}_G \right\rangle
\nonumber \\
&&=
\left\langle \Psi^{T_{QQ}}_G \right| |\mathbf{r}_2 - \mathbf{r}_3| \left| \Psi^{T_{QQ}}_G \right\rangle \,,
\nonumber \\
&&\left\langle \Psi^{T_{QQ}}_G \right| (34)^{-1}(34)|\mathbf{r}_2 - \mathbf{r}_3| (34)^{-1}(34) \left| \Psi^{T_{QQ}}_G \right\rangle
\nonumber \\
&&=
\left\langle \Psi^{T_{QQ}}_G \right| |\mathbf{r}_2 - \mathbf{r}_4| \left| \Psi^{T_{QQ}}_G \right\rangle \,,
\nonumber \\
&&\left\langle \Psi^{T_{QQ}}_G \right| (12)^{-1}(12) |\mathbf{r}_2 - \mathbf{r}_4| (12)^{-1}(12) \left| \Psi^{T_{QQ}}_G \right\rangle
\nonumber \\
&&=
\left\langle \Psi^{T_{QQ}}_G \right| |\mathbf{r}_1 - \mathbf{r}_4| \left| \Psi^{T_{QQ}}_G \right\rangle \,.
\end{eqnarray}
For $T_{QQ'}$,
\begin{eqnarray}
&&\left\langle \Psi^{T_{QQ'}}_G \right| (12)^{-1}(12) |\mathbf{r}_1 - \mathbf{r}_3| (12)^{-1}(12) \left| \Psi^{T_{QQ'}}_G \right\rangle
\nonumber \\
&&=
\left\langle \Psi^{T_{QQ'}}_G \right| |\mathbf{r}_2 - \mathbf{r}_3| \left| \Psi^{T_{QQ'}}_G \right\rangle \,,
\nonumber \\
&&\left\langle \Psi^{T_{QQ'}}_G \right| (12)^{-1}(12) |\mathbf{r}_1 - \mathbf{r}_4| (12)^{-1}(12) \left| \Psi^{T_{QQ'}}_G \right\rangle
\nonumber \\
&&=
\left\langle \Psi^{T_{QQ'}}_G \right| |\mathbf{r}_2 - \mathbf{r}_4| \left| \Psi^{T_{QQ'}}_G \right\rangle \,.
\end{eqnarray}
For $us\bar{b}\bar{b}$,
\begin{eqnarray}
&&\left\langle \Psi^{us\bar{b}\bar{b}}_G \right| (34)^{-1}(34) |\mathbf{r}_1 - \mathbf{r}_3| (34)^{-1}(34) \left| \Psi^{us\bar{b}\bar{b}}_G \right\rangle
\nonumber \\
&&=
\left\langle \Psi^{us\bar{b}\bar{b}}_G \right| |\mathbf{r}_1 - \mathbf{r}_4| \left| \Psi^{us\bar{b}\bar{b}}_G \right\rangle \,,
\nonumber \\
&&\left\langle \Psi^{us\bar{b}\bar{b}}_G \right| (34)^{-1}(34) |\mathbf{r}_2 - \mathbf{r}_3| (34)^{-1}(34) \left| \Psi^{us\bar{b}\bar{b}}_G \right\rangle
\nonumber \\
&&=
\left\langle \Psi^{us\bar{b}\bar{b}}_G \right| |\mathbf{r}_2 - \mathbf{r}_4| \left| \Psi^{us\bar{b}\bar{b}}_G \right\rangle \,.
\nonumber \\
\end{eqnarray}
Therefore, the relative distances reflect the symmetry property.
%%%%%%%%%%%%%%%%%%%%		Figures of Quark Distribution for Tetraquark		%%%%%%%%%%%%%%%%%%%%
\begin{figure}[t]
	\centering
	
	\includegraphics[width=\columnwidth]{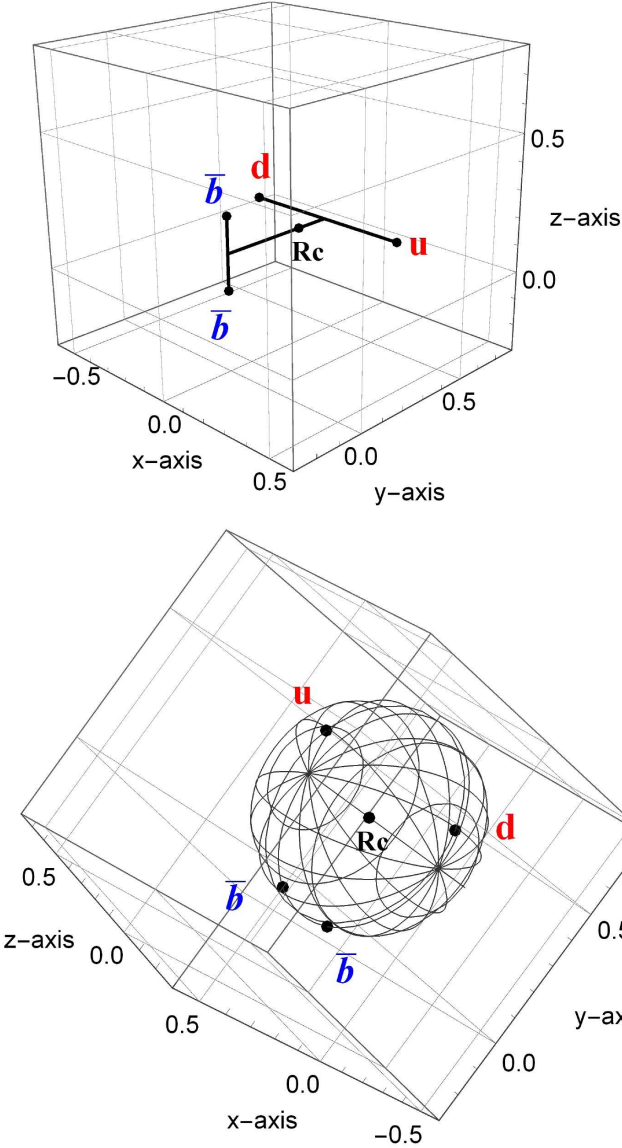}
	\caption{The spatial distributions of the quarks for $T_{bb}$ in fm unit. $\mathbf{R_c}$ is the geometric center of the four quarks (the center of the sphere). The quark positions are the same in both figures. In the bottom figure, the diameter of the sphere is 0.725 fm.}
    \label{TbbSize}
\end{figure}
\begin{figure}[t]
	\centering
	
	\includegraphics[width=\columnwidth]{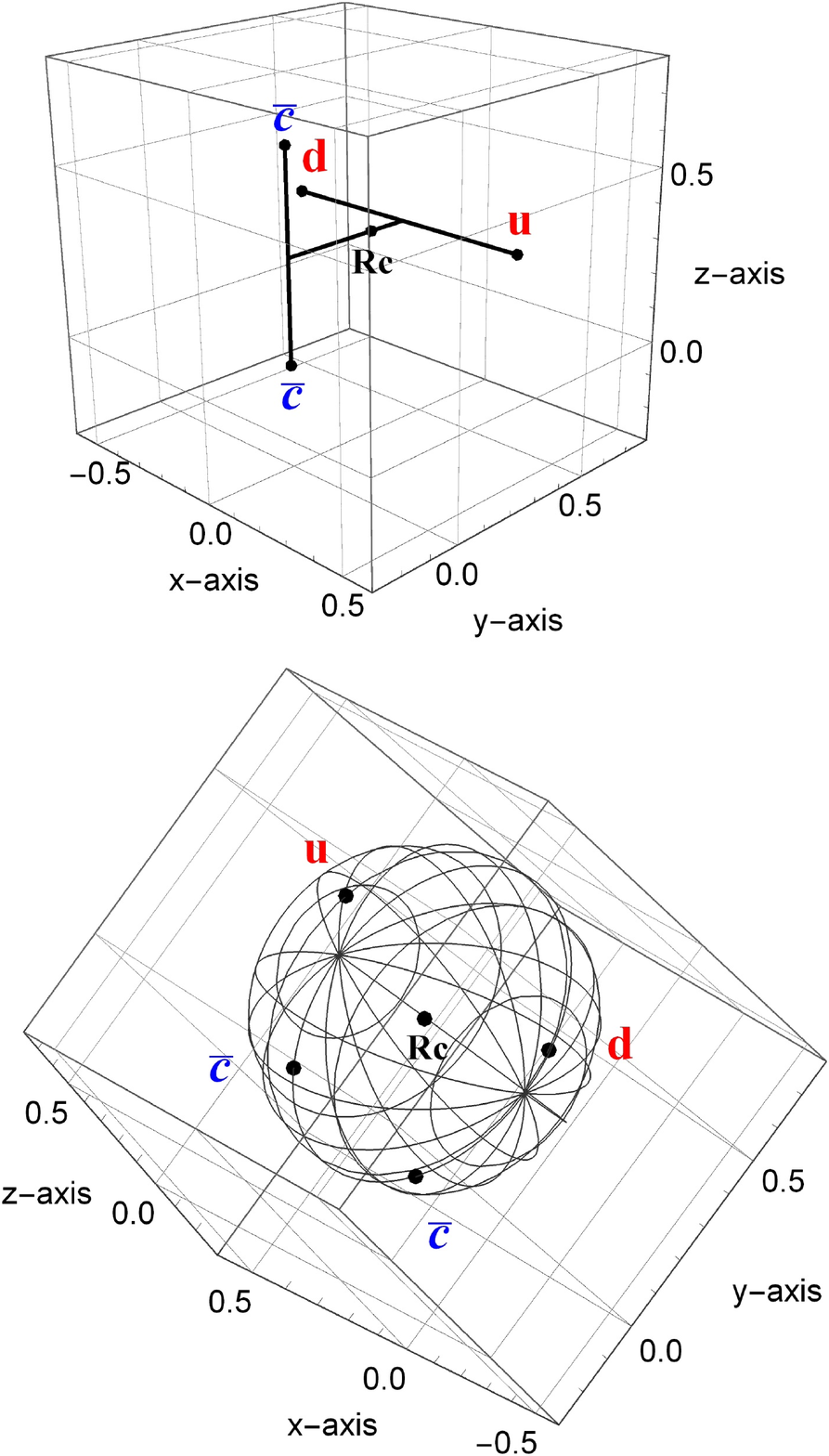}
	\caption{Same as Figure~\ref{TbbSize} but for $T_{cc}$. The diameter of the sphere is 0.866 fm.}
    \label{TccSize}
\end{figure}
\begin{figure}[t]
	\centering
	
	\includegraphics[width=\columnwidth]{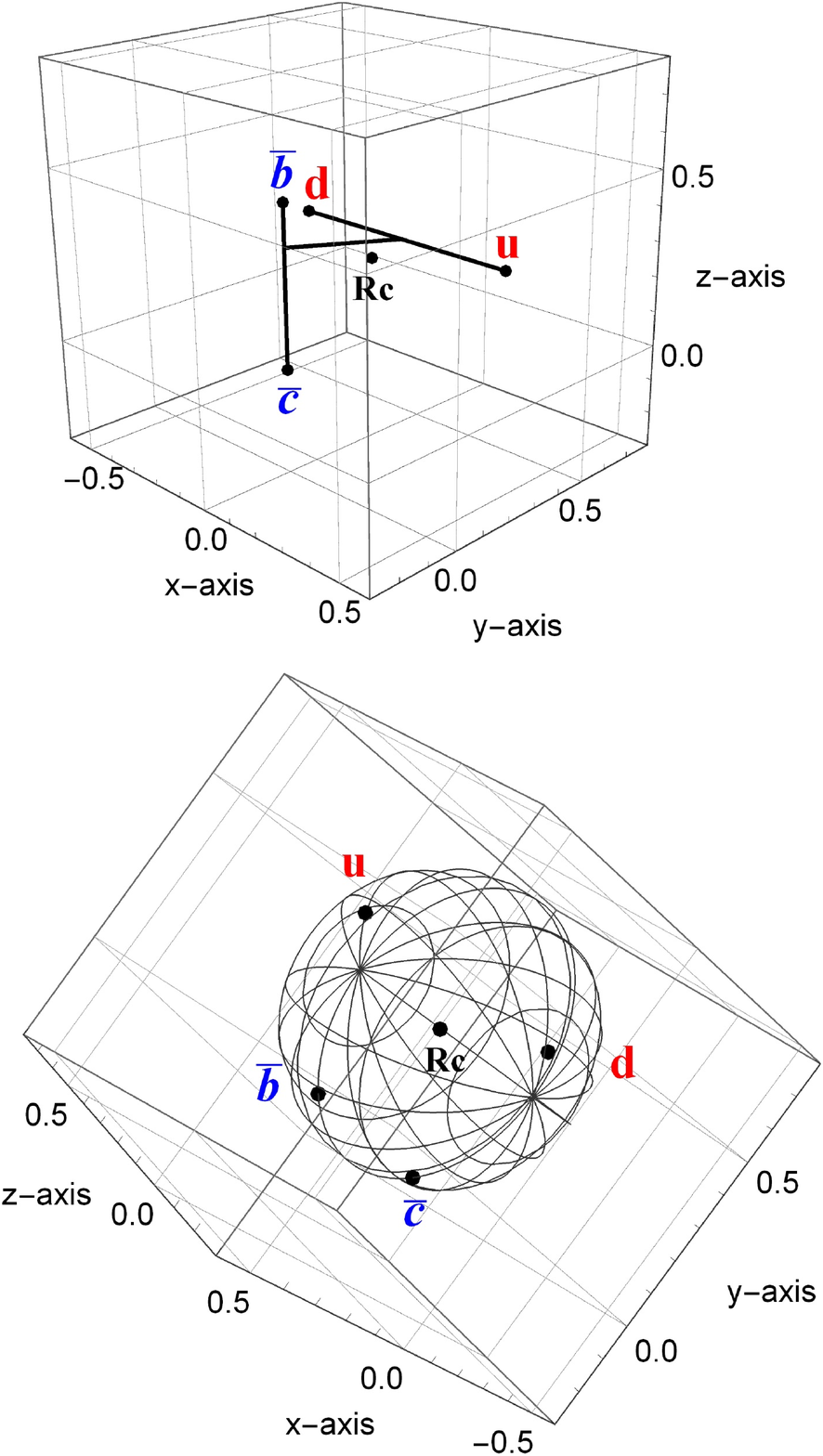}
	\caption{Same as Figure~\ref{TbbSize} but for $T_{cb}$. The diameter of the sphere is 0.789 fm.}
    \label{TcbSize}
\end{figure}
\begin{figure}[t]
	\centering
	
	\includegraphics[width=\columnwidth]{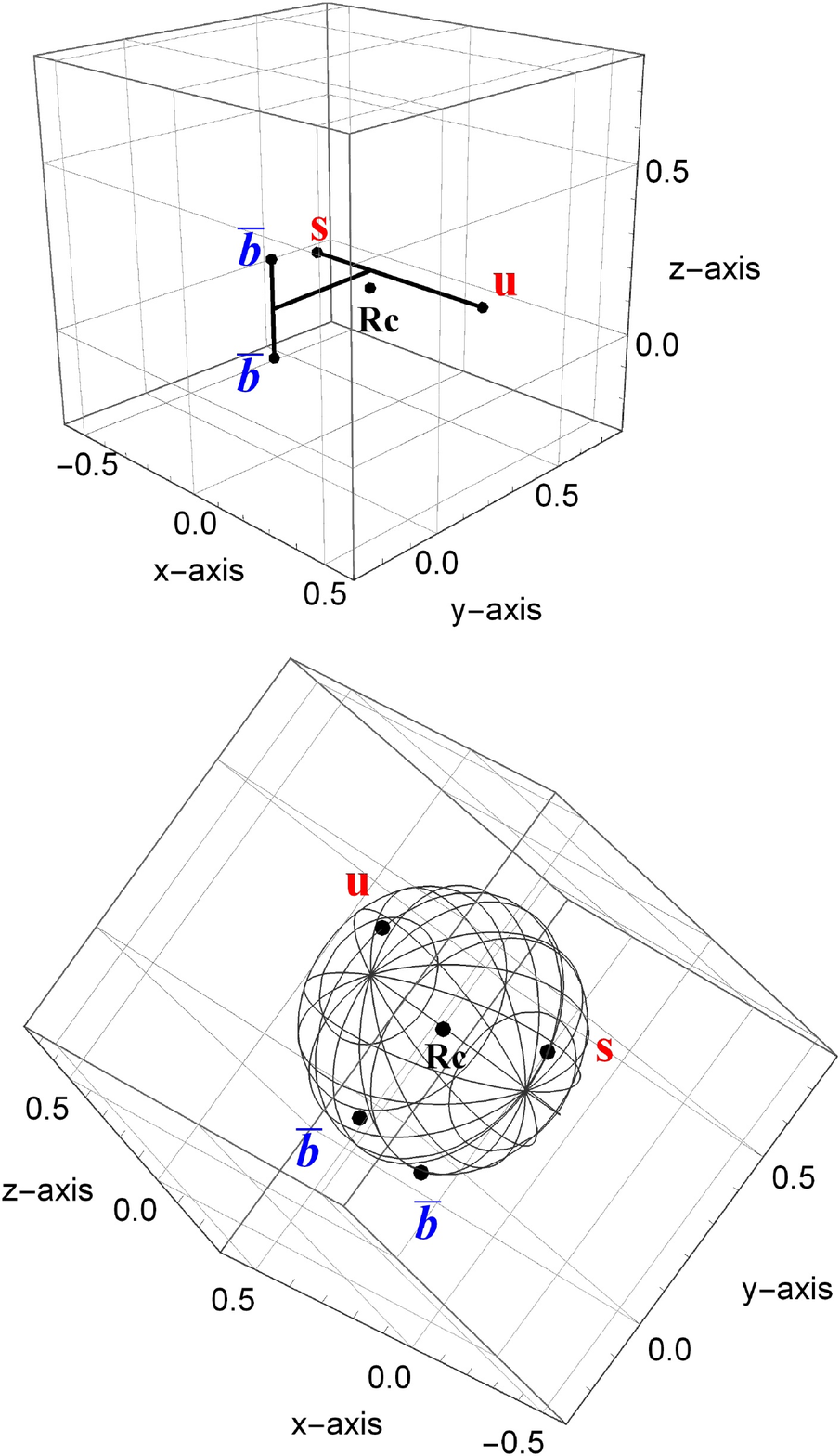}
	\caption{Same as Figure~\ref{TbbSize} but for $us\bar{b}\bar{b}$. The diameter of the sphere is 0.677 fm.}
    \label{usbbSize}
\end{figure}
%%%%%%%%%%%%%%%%%%%%%%%%%%%%%%%%%%%%%%%%%%%%%%%%%%%%%%%%%%%%%%%%%%%%%%%%%%%%%%%%%%%%%%%%%%%%%%

On the other hand, without loss of generality, one can place one of the quarks at the origin of the Cartesian system, and also the second quark can be located on one of the axes. Then the degrees of freedom for the tetraquark system is in general 7. By the symmetry property, the degrees of freedom reduces to 3 for $T_{QQ}$ or 4 for $T_{QQ'}$ and $us\bar{b}\bar{b}$.
Now we can prove for $T_{QQ}$ structure that the three independent vectors, $\mathbf{R}_{(1,2)} \equiv (\mathbf{r}_1 - \mathbf{r}_2)$, $\mathbf{R}_{(3,4)} \equiv (\mathbf{r}_3 - \mathbf{r}_4)$, and $\mathbf{R}' \equiv \frac{1}{2} (\mathbf{r}_1 + \mathbf{r}_2 - \mathbf{r}_3 - \mathbf{r}_4)$, which correspond to the Jacobi coordinates $\mathbf{x}_1$, $\mathbf{x}_2$, and $\mathbf{x}_3$ are orthogonal to each other.
\begin{eqnarray}
&&\left\langle \Psi^{T_{QQ}}_G \right| \left( \mathbf{R}_{(1,2)} \cdot \mathbf{R}_{(3,4)} \right) \left| \Psi^{T_{QQ}}_G \right\rangle
\nonumber \\
&&=
\left\langle \Psi^{T_{QQ}}_G \right| (12)^{-1} (12) \left( \mathbf{R}_{(1,2)} \cdot \mathbf{R}_{(3,4)} \right) (12)^{-1} (12) \left| \Psi^{T_{QQ}}_G \right\rangle
\nonumber \\
&&=
-\left\langle \Psi^{T_{QQ}}_G \right| \left( \mathbf{R}_{(1,2)} \cdot \mathbf{R}_{(3,4)} \right) \left| \Psi^{T_{QQ}}_G \right\rangle=0 \,,
\end{eqnarray}
\begin{eqnarray}
&&\left\langle \Psi^{T_{QQ}}_G \right| \left( \mathbf{R}_{(1,2)} \cdot \mathbf{R}' \right) \left| \Psi^{T_{QQ}}_G \right\rangle
\nonumber \\
&&=
\left\langle \Psi^{T_{QQ}}_G \right| (12)^{-1} (12) \left( \mathbf{R}_{(1,2)} \cdot \mathbf{R}' \right) (12)^{-1} (12) \left| \Psi^{T_{QQ}}_G \right\rangle
\nonumber \\
&&=
-\left\langle \Psi^{T_{QQ}}_G \right| \left( \mathbf{R}_{(1,2)} \cdot \mathbf{R}' \right) \left| \Psi^{T_{QQ}}_G \right\rangle =0 \,,
\end{eqnarray}
\begin{eqnarray}
&&\left\langle \Psi^{T_{QQ}}_G \right| \left( \mathbf{R}_{(3,4)} \cdot \mathbf{R}' \right) \left| \Psi^{T_{QQ}}_G \right\rangle
\nonumber \\
&&=
\left\langle \Psi^{T_{QQ}}_G \right| (34)^{-1} (34) \left( \mathbf{R}_{(3,4)} \cdot \mathbf{R}' \right) (34)^{-1} (34) \left| \Psi^{T_{QQ}}_G \right\rangle
\nonumber \\
&&=
-\left\langle \Psi^{T_{QQ}}_G \right| \left( \mathbf{R}_{(3,4)} \cdot \mathbf{R}' \right) \left| \Psi^{T_{QQ}}_G \right\rangle =0 \,.
\end{eqnarray}
These relations show that we can describe the positions of the four quarks in $T_{cc}$ or $T_{bb}$ with the three independent vectors as shown in Figures~\ref{TbbSize}, \ref{TccSize}. 

For $T_{cb}$, $\mathbf{R}_{(1,2)}$ and $\mathbf{R}_{(3,4)}$ can be taken as two orthogonal vectors, but for the third vector $\left\langle\Psi^{T_{cb}}_G\right|\left(\mathbf{R}_{(3,4)}~\cdot~\mathbf{R}'\right)\left|\Psi^{T_{cb}}_G\right\rangle\neq0$ although $\left\langle\Psi^{T_{cb}}_G\right|\left(\mathbf{R}_{(1,2)}~\cdot~\mathbf{R}'\right)\left|\Psi^{T_{cb}}_G\right\rangle = 0$. 
However,  
$\mathbf{R}'= \left( \mathbf{r}_1 + \mathbf{r}_2 \right)/2 - \left( m_3 \mathbf{r}_3 + m_4 \mathbf{r}_4 \right)/\left( m_3+m_4 \right)$ and linearly independent from  $\mathbf{R}_{(3,4)}$ and thus span the linearly independent third direction as shown in Figure~\ref{TcbSize}. 
One can also show that the non-vanishing component along the $\mathbf{R}_{(3,4)}$ direction for $T_{cb}$ and $\mathbf{R}_{(1,2)}$ direction for $us\bar{b}\bar{b}$ in the wave functions are described through the excited orbital states that produce the asymmetry in the configuration of the Jacobi coordinates set 1 as can be seen in Figures~\ref{TcbSize} and \ref{usbbSize}, respectively. Specifically, since the spatial bases with $(l_1,l_2,l_3)=(0,0,0)$ are of even powers with respect to the Jacobi coordinates $\mathbf{x}_1$, $\mathbf{x}_2$, and $\mathbf{x}_3$, one can find that the spatial part integration of $\left\langle\Psi^{T_{cb}}_G\right|\left(\mathbf{R}_{(3,4)}~\cdot~\mathbf{R}'\right)\left|\Psi^{T_{cb}}_G\right\rangle$ becomes zero if considering only the spatial bases with $(l_1,l_2,l_3)=(0,0,0)$. This implies that the calculations only with the spatial basis $\psi^{Spatial}_{[0,0,0,0,0,0]}$ lead to the $T_{QQ}$ like structure even for both $T_{cb}$ and $us\bar{b}\bar{b}$, which is far away from the real structures in nature. 

In addition, using the relations of the relative distances in Table~\ref{RelDistances}, it is possible to specify the spatial positions of the quarks in the tetraquark structure as the points on the surface of a sphere. The results are depicted in Figures~\ref{TbbSize}-\ref{usbbSize} with the center specified as $\mathbf{R_c}$ in the same scales.

%%%%%%%%%%%%%%%%%%%%%%%%%%%%%%%%%%%%%%%%%%%%%%%%%%%%%%%%%%%%%%%%%%%%%%%%%%%%%%%%%%%%%%%%%%%%%%%
\subsection{Tetraquarks in a Baryon Structure}
\label{BaryonlikeSec}
%%%%%%%%%%%%%%%%%%%%%%%%%%%%%%%%%%%%%%%%%%%%%%%%%%%%%%%%%%%%%%%%%%%%%%%%%%%%%%%%%%%%%%%%%%%%%%%

The structure ($QQ\bar{q}\bar{q}$) is expected to be similar to that of an antibaryon $\bar{Q}_1 \bar{q} \bar{q}$ with the heavy antiquark $\bar{Q}_1$ replaced by the diquark $(QQ)$\cite{Silvestre:ZPC1986}. However, in $qqQ_1$ structure, $Q_1$ is in fact composed of two heavy antiquarks so that it can be of either color [$\mathbf{3}$] or [$\bar{\mathbf{6}}$]. For an antiquark pair of color [$\bar{\mathbf{6}}$], it cannot be regarded as a point particle because, at short distance, the Coulomb potential gives strong repulsion due to the fact that the color matrix element $\langle \boldsymbol{\lambda}^c_i \boldsymbol{\lambda}^c_j \rangle$ is $4/3$ for the color [$\bar{\mathbf{6}}$] while it is $-8/3$ for the color [$\mathbf{3}$]. Thus, in terms of color structure, this means that the $\left( q_1 q_2 \right)^{\mathbf{\bar{3}}} \otimes \left( \bar{q}_3 \bar{q}_4 \right)^{\mathbf{3}}$ channel, which is the only color configuration in $qqQ_1$, dominates the $\left( q_1 q_2 \right)^{\mathbf{6}} \otimes \left( \bar{q}_3 \bar{q}_4 \right)^{\mathbf{\bar{6}}}$ channel in $T_{QQ'}$ tetraquark structure. 
This assumption can be tested by computing the antidiquark $(\bar{Q}\bar{Q'})$ mass in the color $[\mathbf{3}]$ state, and then putting it into the evaluation of the baryon-like mass ($q$-$q'$-$Q_1$) where $Q_1$ replaces the antidiquark $(\bar{Q}\bar{Q'})$.
An isolated diquark is in principle an ill defined concept as its non-zero color makes it a gauge dependent quantity so that one can add any other gauge dependent gluon field configuration to change its mass. What that means is that inside a tetraquark or baryon we are free to include any fractional amount of the interaction between the diquark and other color source to define the diquark mass. For example, the division of the interaction terms between light quark and heavy quark pairs in a tetraquark given in Table \ref{TetraSeparation} is in principle arbitrary. Furthermore, the different spin structure between a heavy quark and diquark, which can be of either spin 1 or 0, could induce different spin interaction for the two different configurations depending on the quantum numbers. 
Thus, we define the diquark mass in a constituent quark model to be the sum of the masses, the interactions between the two antiquarks, and their relative kinetic term all in an isolated color $[\mathbf{3}]$ and spin 0 or 1 configuration.
We can then calculate the hypothetical baryon-like mass assuming a ($q$-$q'$-$Q_1$) structure and call it the tetraquark in a baryon structure, where the subscript 1 denotes the spin of the antidiquark. 
As shown in Table~\ref{Baryonlike}, the tetraquark mass is almost reproduced in the case of $T_{bb}$ and the difference becomes larger in the other tetraquark states.
%%%%%%%%%%%%%%%%%%		Table 9. Tetraquarks in Baryon-like Structre		%%%%%%%%%%%%%%%%%%%%%
\begin{table}[h!]

\caption{Masses of tetraquarks ($qq'\bar{Q}\bar{Q'}$) calculated in a baryon-like structure ($q$-$q'$-$Q_1$), where the isolated antidiquark ($\bar{Q}\bar{Q'}$) mass is used for the  $Q_1$ mass. For comparison, the values in Table~\ref{result} are presented in Column 3. All the masses are in MeV unit.}	
\centering
\begin{tabular}{cccccccc}
\hline
\hline
\multirow{2}{*}{Configuration}	&&&&		Baryon 		&&&		Tetraquark 	\\
									&&&&	Structure		&&&		Structure	\\
\hline
				$T_{cc}$			&&&&		3920		&&&		3873	\\
				$T_{cb}$			&&&&		7238		&&&		7212	\\
		$us\bar{b}\bar{b}$		&&&&		10702		&&&		10694	\\
				$T_{bb}$			&&&&		10517		&&&		10517	\\
\hline 
\hline
\label{Baryonlike}
\end{tabular}
\end{table}
%%%%%%%%%%%%%%%%%%%%%%%%%%%%%%%%%%%%%%%%%%%%%%%%%%%%%%%%%%%%%%%%%%%%%%%%%%%%%%%%%%%%%%%%%%%%%%%
The difference is related to how much the color $\left( q_1 q_2 \right)^{\mathbf{\bar{3}}} \otimes \left( \bar{q}_3 \bar{q}_4 \right)^{\mathbf{3}}$ channel dominates over the $\left( q_1 q_2 \right)^{\mathbf{6}} \otimes \left( \bar{q}_3 \bar{q}_4 \right)^{\mathbf{\bar{6}}}$ channel. As shown in Figures~\ref{TccGround}-\ref{usbbGround}, the contribution of the dominant CS basis in the first quanta is much larger than the others in each tetraquark state, tendency of which is more apparent in $T_{bb}$ than the other tetraquarks. Moreover, looking at the sizes of the antidiquarks in Table~\ref{AntiDiquark}, the corresponding relative distances in the tetraquark structure in Table~\ref{RelDistances} are larger. However, it becomes larger only by 0.006 fm in $T_{bb}$ while it is 0.149 fm in $T_{cc}$. This implies that treating the antiquark pair such as ($\bar{b}\bar{b}$) in $T_{bb}$ as an isolated diquark seems a better approximation when the mass becomes heavier.
%%%%%%%%%%%%%%%%%%%%%		Table 10. Anti-diquark masses and sizes		%%%%%%%%%%%%%%%%%%%%%%%%%%%
\begin{table}[h!]

\caption{The masses and sizes of isolated antidiquarks in the color $[\mathbf{3}]$ and the spin $1$ state. Comparing the sizes to those in Table~\ref{RelDistances}, the effect of surrounding charge becomes smaller for heavier quarks.}	
\centering
\begin{tabular}{cccccccc}
\hline
\hline
\multirow{2}{*}{Antidiquark}	&&&	\multirow{2}{*}{Mass(MeV)}	&&	Variational	&&	\multirow{2}{*}{Size(fm)} 	\\
									&&&									&&	Parameter (fm$^{-2}$)	&&	\\
\hline
	$\bar{c}\bar{c}$		&&&		3609		&&	$a=9.5$		&&		0.461	\\
	$\bar{b}\bar{b}$		&&&		10234		&&	$a=30.5$	&&		0.262	\\
	$\bar{c}\bar{b}$		&&&		6947		&&	$a=14.4$	&&		0.378	\\
\hline 
\hline
\label{AntiDiquark}
\end{tabular}
\end{table}
%%%%%%%%%%%%%%%%%%%%%%%%%%%%%%%%%%%%%%%%%%%%%%%%%%%%%%%%%%%%%%%%%%%%%%%%%%%%%%%%%%%%%%%%%%%%%%%

%%%%%%%%%%%%%%%%		Table 11. Relative distances in various configurations		%%%%%%%%%%%%%
\begin{table}[h!]

\caption{The relative distances between the quarks in various configuration. The size of an isolated $ud$ diquark in the color $[\bar{\mathbf{3}}]$ and spin $0$ state is given in Column 2. The distances in the baryon-like ($q$-$q'$-$Q_1$) structure are given in Columns 3-5. The distances in the baryons $\Lambda_c$ and $\Lambda_b$ are given in Columns 6-7.}	
\centering
\begin{tabular}{ccccccccccccc}
\hline
\hline
\multirow{2}{*}{Quark Pair}	&&	\multirow{2}{*}{Diquark}	&&	\multicolumn{5}{c}{($q$-$q'$-$Q_1$) Structure}	&&	\multicolumn{3}{c}{Baryon}	\\
\cline{5-9}\cline{11-13}
			&&	&&	$T_{cc}$	&&	$T_{cb}$	&&	$T_{bb}$	&&	$\Lambda_c$&&	$\Lambda_b$	\\
\hline
	(1,2)	&&0.780	&&	0.664	&&	0.661		&&	0.660		&&	0.668		&&	0.662	\\
	(1,3)	&&	-	&&	0.590	&&	0.578		&&	0.574		&&	0.611		&&	0.582	\\
	(2,3)	&&	-	&&	0.590	&&	0.578		&&	0.574		&&	0.611		&&	0.582	\\
\hline 
\hline
\label{udDiquark}
\end{tabular}
\end{table}
%%%%%%%%%%%%%%%%%%%%%%%%%%%%%%%%%%%%%%%%%%%%%%%%%%%%%%%%%%%%%%%%%%%%%%%%%%%%%%%%%%%%%%%%%%%%%%%

Now we look at the size of $(ud)$ pair in various configurations in Table~\ref{udDiquark}. The size of $(ud)$ pair in the diquark configuration becomes smaller in the baryons $\Lambda_c$ and $\Lambda_b$ due to the interaction with the heavy quark. Comparing the size of ($ud$) pair in $\Lambda_b$ and $\Lambda_c$, one finds that the size of ($ud$) pair becomes smaller when the heavy quark is closer to the ($ud$) pair. This tendency can be seen also in the baron-like ($q$-$q'$-$Q_1$) structure. However, in the tetraquark configuration as can be seen in Table~\ref{RelDistances}, the size of ($ud$) pair becomes smaller in $T_{bb}$ than that in $T_{cc}$ even though the relative distance (1,2)-(3,4) is smaller in $T_{cc}$, which shows an opposite tendency to that in the baryon structure.

Considering the tetraquarks $T_{cc}$ and $T_{bb}$ as a baryon-like ($q$-$q$-$Q_1$) structure, the size of $(ud)$ pair in the baryon-like $T_{cc}$($T_{bb}$) is close to that in the baryon $\Lambda_c$($\Lambda_b$). However, as can be seen in Table~\ref{RelDistances}, the size of $(ud)$ pair in $T_{cc}$ is in fact 0.830 fm, which is not so close to that in the baryon-like $T_{cc}$, and is even larger than the size of $(ud)$ pair in the diquark configuration, while the size of $(ud)$ pair in $T_{bb}$ is close to that in the baryon-like $T_{bb}$. The analysis so far points out that $T_{bb}$ indeed has a similar structure as a baryon, while $T_{cc}$ does not.

As discussed in the work\cite{Woosung:NPA2019}, the total mass obtained with the Hamiltonian in Eq.~(\ref{Hamiltonian}) can be divided into `light quark', `heavy quark', `CS' parts as given in Table~\ref{TetraSeparation}. The constant $-D$ term in the Hamiltonian is divided into each quark by multiplying a factor of 1/2. The relative kinetic energy involving $\mathbf{p}_3$, which corresponds to the relative coordinate $\mathbf{x}_3$ connecting the quark and the antiquark pairs can be divided according to the relative contributions depending on the mass of either the quark pair or the antiquark pair.

Considering only the spatial basis $\psi^{Spatial}_{[0,0,0,0,0,0]}$, the relative distance between the light quarks in $T_{cc}$ is the same with that in $T_{bb}$\cite{Woosung:NPA2019}. Comparing the results between the full calculation and the simplified calculation denoted by 1-Basis in Table~\ref{TetraSeparation}, the hyperfine interaction $\sum V^{CS}(i,j)$ in $T_{cc}$ becomes much stronger than that in $T_{bb}$. This can be understood from the change in $V^{CS}(1,2)$, which implies that the attraction coming from $V^{CS}(1,2)$ partially spreads into the attraction coming from $\sum V^{CS}(i,j)$ in $T_{cc}$. As a result, for $T_{bb}$ as shown in Table~\ref{TetraSeparation}, the attraction from the hyperfine part between $\bar{b}$ and $u$ (or $d$) quarks is just -5.7 MeV, while that in $T_{cc}$ is -69.4 MeV. This is one of the major reasons that the distance between the light quark pair and the heavy antiquark pair is closer in $T_{cc}$ than that in $T_{bb}$ as seen in Table~\ref{RelDistances}. This closer distance in $T_{cc}$ causes the slightly larger size of ($ud$) pair in $T_{cc}$ than in $T_{bb}$.
%%%%%%%%%%%%%%%%%%%%%%%		Table 12. Tbb & Tcc Separations		%%%%%%%%%%%%%%%%%%%%%%%%%%%%%%%
\begin{widetext}

\begin{table}[h!]

\caption{Contributions to the $T_{bb} (ud\bar{b}\bar{b})$ and $T_{cc} (ud\bar{c}\bar{c})$ masses from the present work. $(i,j)$ denotes the $i$ and $j$ quarks, where $i,j=1,2$ label the light quarks, and 3, 4 are for the heavy antiquarks in each configuration.  $\sum V^C(i,j)$ and $\sum V^{CS}(i,j)$ covers pairs ($i,j$) except for (1,2) and (3,4) pairs. $D$ is separately added and not included in $V^C(i,j)$. $m_Q$ is the heavy quark mass, and $m_i'$ are defined in Eq.~(\ref{ReducedM}) for each configuration. $\mathbf{p}_i$ is the relative momentum corresponding to the $i$-th Jacobi coordinate $\mathbf{x}_i$. 1-Basis is the result with only one spatial basis $\psi^{Spatial}_{[0,0,0,0,0,0]}$ and the corresponding dominant CS basis.}	
\centering

\begin{tabular}{ccccccccc}
\hline
\hline
	\multirow{2}{*}{Overall}	&&	\multirow{2}{*}{Contribution}	&& \multicolumn{2}{c}{$T_{bb}$}	&& \multicolumn{2}{c}{$T_{cc}$}	\\
\cline{5-6}\cline{8-9}
				&&		&&Full Calculation&1-Basis&&Full Calculation&1-Basis\\
\hline
		Heavy Quark			&&			2$m_Q$		&&	10674.0	&	10674.0	&&	3844.0	&	3844.0	\\
				&&$\frac{\mathbf{p}^2_2}{2 m'_2}$	&&	206.8	&	220.0	&&	142.5	&	221.8	\\
&&$\frac{m_q}{m_Q+m_q} \frac{\mathbf{p}^2_3}{2 m'_3}$&&16.4	&	15.3	&&	53.8	&	38.0	\\
								&&		$V^C(3,4)$		&&$-188.8$	&$-190.8$	&&	19.3	&	4.2		\\
					&&	$\frac{1}{2} \sum V^C(i,j)$	&&	115.8	&	137.6	&&	159.1	&	168.5	\\
								&&			$-D$		&&$-917.0$	&$-917.0$	&&$-917.0$	&$-917.0$	\\
			Subtotal			&&						&&	9907.2	&	9939.1	&&	3301.8	&	3359.5	\\
\hline
		Light Quark			&&			2$m_q$		&&	684.0	&	684.0	&&	684.0	&	684.0	\\
				&&$\frac{\mathbf{p}^2_1}{2 m'_1}$	&&	494.1	&	495.3	&&	424.1	&	478.2	\\
&&$\frac{m_Q}{m_Q+m_q} \frac{\mathbf{p}^2_3}{2 m'_3}$	&&255.8	&	239.1	&&	302.2	&	213.5	\\
								&&		$V^C(1,2)$		&&	171.3	&	181.6	&&	91.3	&	188.8	\\
					&&	$\frac{1}{2} \sum V^C(i,j)$	&&	115.8	&	137.6	&&	159.1	&	168.5	\\
								&&			$-D$		&&$-917.0$	&$-917.0$	&&$-917.0$	&$-917.0$	\\
			Subtotal			&&						&&	804.0	&	820.6	&&	743.7	&	816.0	\\
\hline
			CS Interaction		&&		$V^{CS}(3,4)$	&&	7.0		&	6.8		&&	5.3		&	9.3		\\
								&&		$V^{CS}(1,2)$	&&$-195.3$	&$-188.1$	&&$-108.6$	&$-182.6$	\\
								&&	$\sum V^{CS}(i,j)$&&$-5.7$	&	0.0		&&$-69.4$	&	0.0		\\
			Subtotal			&&						&&$-194.0$	&$-181.3$	&&$-172.7$	&$-173.3$	\\
\hline
			Total				&&						&&	10517.2&	10578.4	&&	3872.8	&	4002.2	\\
\hline
\hline
\label{TetraSeparation}
\end{tabular}
\end{table}

\end{widetext}
%%%%%%%%%%%%%%%%%%%%%%%%%%%%%%%%%%%%%%%%%%%%%%%%%%%%%%%%%%%%%%%%%%%%%%%%%%%%%%%%%%%%%%%%%%%%%

%%%%%%%%%%%%%%%%%%%%%%%%%%%%%%%%%%%%%%%%%%%%%%%%%%%%%%%%%%%%%%%%%%%%%%%%%%%%%%%%%%%%%%%%%%%%%%%
\subsection{Comparison with Other Models}
%%%%%%%%%%%%%%%%%%%%%%%%%%%%%%%%%%%%%%%%%%%%%%%%%%%%%%%%%%%%%%%%%%%%%%%%%%%%%%%%%%%%%%%%%%%%%%%
We now compare our present results with other works in Ref.~\cite{Karliner:PRL2017} and Ref.~\cite{Silvestre:ZPC1993}. In a simplified constituent quark model\cite{Karliner:PRL2017}, the masses for $T_{cc}$ and $T_{bb}$ are given as
\begin{eqnarray}
M_{T_{cc}} &=& 2m^b_c + B(cc) + 2m^b_q + \frac{a_{cc}}{(m^b_c)^2} - \frac{3 a}{(m^b_q)^2}
\nonumber \\
&=&
3421.0 - 129.0 + 726.0 + 14.2 - 150.0 = 3882.2 \,,
\nonumber\\
M_{T_{bb}} &=& 2m^b_b + B(bb) + 2m^b_q + \frac{a_{bb}}{(m^b_b)^2} - \frac{3 a}{(m^b_q)^2}
\nonumber \\
&=&
10087.0 - 281.4 + 726.0 + 7.8 - 150.0 = 10389.4\,,
\nonumber \\
\label{SimEq}
\end{eqnarray}
where $m^b_{c,b,q}$ are the constituent quark masses of $c$, $b$, and light quark $q$ inside a baryon, and $B(cc)$ ($B(bb)$) is the binding between the charm(bottom) quarks, which can be understood as coming from the extra attraction between two $c(b)$ quarks due to that shorter interquark distance as compared to two light quarks: this attraction can be estimated by studying the quark attractions inside $\Lambda_c$ and $\Xi_{cc}$\cite{Woosung:NPA2019}. $a$'s are multiplicative constants for the CS interaction. Here, $a_{QQ}/(m^b_Q)^2$ is the CS interaction between the two heavy quarks $Q$ corresponding to $V^{CS}(3,4)$ in our model, while $-3a/(m^b_q)^2$ is that between the light quarks $q$ denoted by $V^{CS}(1,2)$ in our model. Then, treating $B(cc)$ ($B(bb)$) as part of the two charm(bottom) quark system, the energy in the simplified model in Eq.~(\ref{SimEq}) can be divided into the charm(bottom) quark, light quark, and CS interaction parts.
The subtotal values in Table~\ref{TetraSeparation} can be regarded as the constituent quark masses in the simplified quark model in Eq.~(\ref{SimEq}). The additional attraction for the heavy quark pairs with respect to the light quark pair in the simple model\cite{Karliner:PRL2017} can be seen from $V^C(3,4)$ being much more attractive than $V^C(1,2)$. 
While the importance of the large additional attraction for heavier quarks denoted by $B(QQ)$ remains a valid point, detailed mass and size differences change other parameters such as the effective quark masses in Eq.~(\ref{SimEq}) so that a simple parametrization formula given in Eq.~(\ref{SimEq}) for the tetraquark masses might become problematic. Furthermore, while Eq.~(\ref{SimEq}) does not allow for the color spin interaction between light and heavy quarks, our full calculations show the presence of such terms as given in Table XII, which becomes $-5.7$ and $-69.4$ MeV for $T_{bb}$ and $T_{cc}$, respectively. Also, the discrepancy becomes larger for $T_{bb}$ suggesting that nonlinear quark mass dependence becomes also important. Therefore, care should be taken when simple parametrizations that work for normal hadrons are generalized to more complicated configurations.

There is another work\cite{Silvestre:ZPC1993} using the complete set of harmonic oscillator bases, which is similar to ours but different in part as  we introduced the rescaled form of the harmonic oscillator bases. The hamiltonian in their model is as follows.
\begin{eqnarray}
H
=
\sum_i \left(m_i + \frac{\mathbf{p}_i^2}{2 m_i} \right) + \sum_{i<j} V_{ij}(r_{ij}) \,,
\end{eqnarray}
with
\begin{eqnarray}
\hspace{-2cm}V_{ij}(r_{ij})
&=&
- \frac{3}{16} \lambda_i \lambda_j \bigg[ - \frac{\kappa}{r_{ij}} + \frac{r_{ij}}{a_0^2} - D
\nonumber \\
&&\hspace{2cm}+ \frac{\hbar^2 c^2 \kappa}{m_i m_j c^4} \frac{e^{-r_{ij}/r_0}}{r_0^2 r_{ij}} \mathbf{\sigma}_i \mathbf{\sigma}_j \bigg] \,,
\end{eqnarray}
where the hyperfine interaction is of $\exp[-r_{ij}]$, instead of $\exp[-r^2_{ij}]$ in our model. Also, they fixed the parameters $\kappa$ and $(r_0)_{ij}$ by fitting them to the experimental values, while the hyperfine part in our model has additional mass dependence appearing as in Eqs. (\ref{Parameter1})-(\ref{Parameter2}). The fitting parameters in their model are as follows.
\begin{eqnarray}
&\kappa=102.67 \, \textrm{MeV fm}, \quad a_0=0.0326 \, \textrm{(MeV$^{-1}$fm)$^{1/2}$},& \nonumber \\
&D=913.5  \, \textrm{MeV}, \quad r_0=0.4545 \, \textrm{fm},& \nonumber \\
&m_{u}=337 \, \textrm{MeV}, \qquad m_{s}=600 \, \textrm{MeV}, &\nonumber \\
&m_{c}=1870 \, \textrm{MeV}, \qquad m_{b}=5259 \, \textrm{MeV}.	&
\label{SilvestreFit}
\end{eqnarray}
Comparing with our results in Table~\ref{ComparisonSilvestre}, the two independent models give almost the same binding energies for each tetraquark state. However, there is a large difference in the mass of $T_{cc}$, which is due to the larger obtained masses for the $D$ and $D^*$ mesons in their model, which were found to be 1891 MeV and 2021 MeV, respectively. Compared to the total mass of the $D$ and $D^*$ mesons in our model, the difference is 52 MeV, which approximately accounts for the difference in the mass of $T_{cc}$ between the two models.
Also, in their model, the lowest threshold for $us\bar{b}\bar{b}$ is $B$ $B_s^*$, while it is $B^*$ $B_s$ in our model as in the experiment. In the experiment, the total mass of the $B$ and $B_s^*$ mesons is slightly larger than that of the $B^*$ and $B_s$ mesons by approximately 2.7 MeV, while the corresponding difference in our model calculation is $4.7$ MeV. 
%%%%%%%%%%%%%%%%%%%%%%%%		Table 14. Comparison with Silvestre		%%%%%%%%%%%%%%%%%%%%%%%%%
\begin{table}[h!]
\caption{The masses and the binding energies $B_{T}$ of the tetraquark states obtained in the present work and in Ref.~\cite{Silvestre:ZPC1993}. The masses and $B_{T}$ are in MeV unit.}
\begin{tabular}{cccccccccccc}
\hline
\hline
\multirow{2}{*}{Type}	&&\multirow{2}{*}{($I$, $S$)}&&	\multicolumn{3}{c}{Present Work}&&&	\multicolumn{3}{c}{Ref.~\cite{Silvestre:ZPC1993}}
\\
\cline{5-7}\cline{10-12}
					&&				&&	Mass	&&	$B_{T}$		&&&	Mass	&&	$B_{T}$		\\
\hline 
%%% Tbb %%%%%
$ud\bar{b}\bar{b}$	&&	(0,1)	&&	10517	&&	$-145$	&&&	10525	&&	$-131$	\\
%%% Tcc %%%%%
$ud\bar{c}\bar{c}$	&&	(0,1)	&&	3873	&&	+13		&&&	3931	&&	+19		\\
%%% Tcb %%%%%
$ud\bar{c}\bar{b}$	&&	(0,1)	&&	7212	&&	$-3$	&&&	7244	&&	+1		\\
%%% usbb %%%%
$us\bar{b}\bar{b}$	&&	(1/2,1)	&&	10694	&&	$-42$	&&&	10680	&&	$-40$ 	\\
\hline 
\hline
\label{ComparisonSilvestre}
\end{tabular}
\end{table}
%%%%%%%%%%%%%%%%%%%%%%%%%%%%%%%%%%%%%%%%%%%%%%%%%%%%%%%%%%%%%%%%%%%%%%%%%%%%%%%%%%%%%%%%%%%%%

%%%%%%%%%%%%%%%%%%%%%%%%%%%%%%%%%%%%%%%%%%%%%%%%%%%%%%%%%%%%%%%%%%%%%%%%%%%%%%%%%%%%%%%%%%%%%%%
\section{Summary and Discussion}
%%%%%%%%%%%%%%%%%%%%%%%%%%%%%%%%%%%%%%%%%%%%%%%%%%%%%%%%%%%%%%%%%%%%%%%%%%%%%%%%%%%%%%%%%%%%%%%
We have improved our constituent quark model by introducing the complete set of 3-dimensional harmonic oscillator bases. We also assessed the validity by comparing the ground state wave function for the meson structure with the exact solution of the hydrogen model. The effect turns out to lower the binding energies for the tetraquark systems. In particular, the harmonic oscillator bases for the excited orbital states play a crucial role in obtaining the exact ground state for the tetraquark systems. We have also successfully fitted the parameters in the Hamiltonian in Eq.~(\ref{Hamiltonian}) to most of the observed mesons and baryons allowed in our model. The results are summarized in Table~\ref{result}. Also, from the relative distances between the quarks given in Table~\ref{RelDistances}, we have described the spatial distributions of the quarks in the tetraquark structure in Figures~\ref{TbbSize}-\ref{usbbSize}. By comparing with our earlier work shown in Table~\ref{result}, and the discussions in Sections~\ref{SizeSec}, \ref{BaryonlikeSec}, we have found that a simple Gaussian spatial function fails to provide precise information on the stability and the structure for the tetraquarks so that the detailed treatment as presented in this work should be performed.  Also, by comparing with other work, fitting procedure is important to evaluate the exact values for the masses. One can conclude that while simplified constituent quark model based on universal constants that does not depend on specific configuration and/or simple models based on universal diquarks are intuitively important to identify possible attractive configurations, a detailed full constituent quark model calculation is needed to assess the stability and existence of a compact exotic multiquark configuration.

%%%%%%%%%%%%%%%%%%%%%%%%%%%%%%%%%%%%%%%%%%%%%%%%%%%%%%%%%%%%%%%%%%%%%%%%%%%%%%%%%%%%%%%%%%%%%%%
\section*{Acknowledgments}
This work was supported by Samsung Science and Technology
Foundation under Project Number SSTF-BA1901-04, and by the Korea National Research Foundation under the grant number 2017R1D1A1B03028419(NRF).
%
%
%
%

%%%%%%%%%%%%%%%%%%%%%%%%%%%%%%%%%%%%%%%%%%%%%%%%%%%%%%%%%%%%%%%%%%%%%%%%%%%%%%%%%%%%%%%%%%%%%%%
%
%
%												Appendix
%
%
%%%%%%%%%%%%%%%%%%%%%%%%%%%%%%%%%%%%%%%%%%%%%%%%%%%%%%%%%%%%%%%%%%%%%%%%%%%%%%%%%%%%%%%%%%%%%%%
\section*{Appendix}
\numberwithin{equation}{subsection}
%%%%%%%%%%%%%%%%%%%%%%%%%%%%%%%%%%%%%%%%%%%%%%%%%%%%%%%%%%%%%%%%%%%%%%%%%%%%%%%%%%%%%%%%%%%%%%%
\subsection{Harmonic Oscillator Bases in Mesons}
%%%%%%%%%%%%%%%%%%%%%%%%%%%%%%%%%%%%%%%%%%%%%%%%%%%%%%%%%%%%%%%%%%%%%%%%%%%%%%%%%%%%%%%%%%%%%%%

To construct the spatial function for the meson structure, we solve the Schr$\ddot{\rm o}$dinger equation for the 3-dimensional symmetric harmonic oscillator. In the spherical coordinates system, the wave function can be separated into the radial part and the angular part such as $\psi(r, \theta, \phi) = R(r) Y^m_l (\theta, \phi)$. The solution of the angular part is known to be the spherical harmonics. For the radial part of the equation, the solution is obtained in terms of the associated Laguerre polynomials. The orthonormalized radial part wave function is known as follows.
\begin{eqnarray}
R_{n,l}(r) = \sqrt{\frac{2 \, \Gamma (n+1)}{\Gamma \left( n+l+\frac{3}{2} \right)}} \, r^l \exp \left[-\frac{r^2}{2} \right] L^{l+\frac{1}{2}}_n \left( r^2 \right). \qquad \quad
\label{RadialWave1}
\end{eqnarray}
where $L^{l+\frac{1}{2}}_n (r^2)$ is the associated Laguerre polynomials. For our purpose of introducing the harmonic oscillator bases to our model, we have modifed Eq.~(\ref{RadialWave1}) by rescaling the radial distance $r$ to $\sqrt{2 a} \, x$, where $x$ is the magnitude of the Jacobi coordinate $\mathbf{x}$, connecting the quark and the antiquark in the meson structure, and $a$ is the variational parameter corresponding to the coordinate $\mathbf{x}$. 
Then the spatial part of the total wave function is constructed by combining the spherical harmonics as follows.
\begin{eqnarray}
\hspace{-0.9cm}\psi^{Spatial}_{[n,l, m]}(\mathbf{x}) = \psi(x, \theta, \phi)^{Spatial}_{[n,l, m]} = R_{n,l}(x) Y^m_l (\theta, \phi) \,.
\label{MesonSpatial}
\end{eqnarray}
where $R_{n,l}(x)$ is the rescaled radial part wave function. Here, the quantum numbers $n$, $l$ indicate the principal quantum number, and the orbital angular momentum, respectively. For mesons in the $l=0$ state, each of the harmonic oscillator bases should be of $l=0$, and they compose the spatial part of the wave function in the meson structure.

On the other hand, the permutation symmetry of the spatial bases depends on the power of the Jacobi coordinate $|\mathbf{x}|$, which is contained only in the radial part of the spatial function. From Eq.~(\ref{RadialWave1}), it is recognized that the permutation symmetry is determined by the angular momentum quantum number $l$. In our case, for all the bases in the calculations, the angular momentum is $l=0$. Therefore, for the mesons in the ground state, all the spatial bases are symmetric under the permutation (12).

\begin{figure}[h!]
	\centering
	
	\includegraphics[width=\columnwidth]{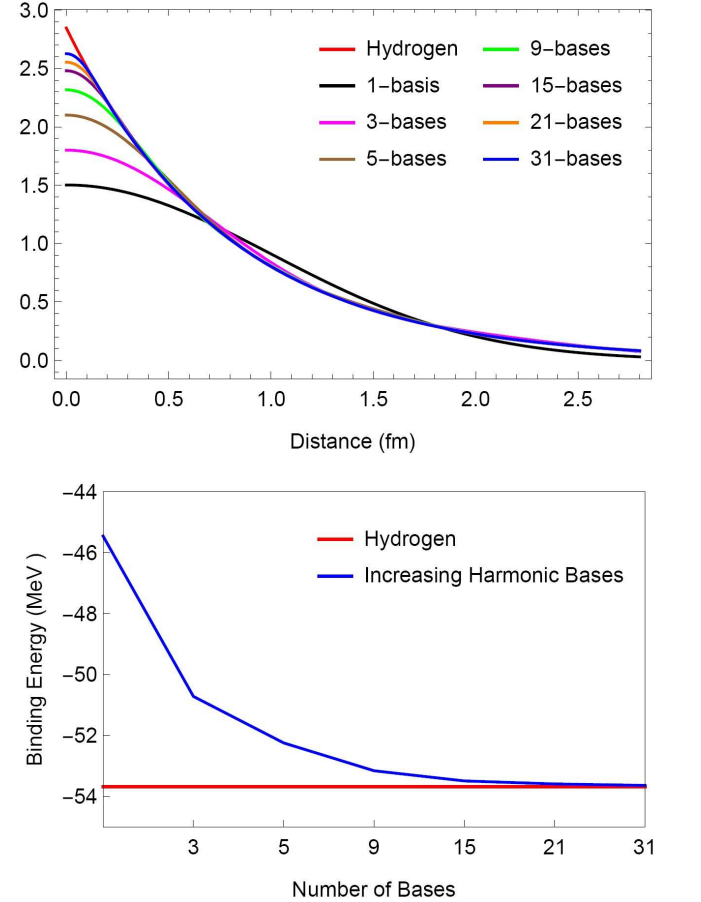}
	\caption{Comparisons between the exact solution and the guasian basis for the  $D$ meson with coulomb potential. In the figures, Hydrogen indicates the exact solutions . The upper figure represents the radial parts of the ground state wave functions as the number of the harmonic oscillator bases increases.  The bottom figure represents the convergence behavior of the binding energy as the number of bases increases.  The binding energy in the hydrogen model is -53.69 MeV, and the binding energy with the 31 harmonic oscillator bases is -53.64 MeV, which is very close to the value in the hydrogen model.}
    \label{Dmeson}
\end{figure}

To assess the validity of the harmonic oscillator bases approach, we compared the meson structure to the hydrogen atom in hadron picture. To do this, we considered only the kinetic energy and the Coulomb potential in the Hamiltonian of Eq.~(\ref{Hamiltonian}). Then the Hamiltonian reduces to the following form with the kinetic energy in the center of mass frame.
\begin{eqnarray}
H &=&\frac{{\mathbf p}^{2}_{\mathbf{x}}}{2 m'} -\frac{3}{4} \frac{\lambda^{c}_{1}}{2} \,\, \frac{\lambda^{c}_{2}}{2} \left( - \frac{\kappa}{r_{12}}  \right). \qquad
\label{Hamiltonian-meson}
\end{eqnarray}
where $m' = (2 m_1 m_2)/(m_1 + m_2)$, and ${\mathbf p}_{\mathbf{x}}$ is the relative momentum between the quark and the antiquark. We take $m_1=m_c, m_2=m_q$, and the values in Eq.~(\ref{FitParameters}) for the parameters $m_c$, $m_q$, and $\kappa$. For our purpose of the comparison to our model, we also modify the exact solution of the hydrogen atom as that of the relative coordinate $\mathbf{x}$. Then, in hadron picture, the radial part of the ground state wave function for the hydrogen, $R^{Hydrogen}_{[0,0]}(x)$, becomes as follows.
\begin{eqnarray}
R^{Hydrogen}_{[0,0]}(x)
=
2 \left( \frac{\sqrt{2} \kappa}{\hbar^2} \mu \right)^{3/2} \exp [ - a|\mathbf{x}| ] \,.
\label{Hydrogen}
\end{eqnarray}
where $\mu$ is the reduced mass of the system in MeV unit.
In comparing, we choose the $D$ meson as a target, then the reduced mass $\mu$ in Eq.~(\ref{Hydrogen}) becomes $\frac{m_u m_c}{m_u + m_c}$. The results are shown in Figure~\ref{Dmeson}. It is obvious from Figure~\ref{Dmeson} that the more harmonic oscillator bases are included, the more precisely it describes the actual ground state of the meson structure.
We have included the harmonic oscillator bases up to $n=4$, which is enough to obtain the convergent values for the ground state masses of the tetraquarks of interest in this work.

%%%%%%%%%%%%%%%%%%%%%%%%%%%%%%%%%%%%%%%%%%%%%%%%%%%%%%%%%%%%%%%%%%%%%%%%%%%%%%%%%%%%%%%%%%%%%%%
\subsection{Harmonic Oscillator Bases in Baryons}
%%%%%%%%%%%%%%%%%%%%%%%%%%%%%%%%%%%%%%%%%%%%%%%%%%%%%%%%%%%%%%%%%%%%%%%%%%%%%%%%%%%%%%%%%%%%%%%

In baryons, there arise additional degrees of freedom due to the second Jacobi coordinate $\mathbf{x}_2$. Furthermore, there are  contributions from combinations of nonzero internal relative orbital angular momenta, satisfying zero total orbital angular momentum of the ground state baryon structure. Using a similar method as in the meson structure, we construct the complete set of orthonormalized harmonic oscillator bases.
\begin{eqnarray}
&&\hspace{-0.2cm}\psi(\mathbf{x}_1, \mathbf{x}_2)^{Spatial}_{[n_1,n_2,l_1,1_2]}
\nonumber \\
&&\hspace{-0.1cm}= \sum_{m_1,m_2} C(l_1,m_1,l_2,m_2;l=0,m=0)
\nonumber \\
&&\hspace{0.8cm} \times \, R_{n_1,l_1}(x_1) R_{n_2,1_2}(x_2) Y^{m_1}_{l_1} (\theta_1, \phi_1) Y^{m_2}_{l_2} (\theta_2, \phi_2) \,.
\nonumber \\
\end{eqnarray}
where $R_{n_i,l_i}(x_i)$ has the same form as in Eq.~(\ref{MesonSpatial}), and $Y^{m_i}_{l_i} (\theta_i, \phi_i)$ is the spherical harmonic function for the angular part of the $i$-th Jacobi coordinate $\mathbf{x}_i$. $C(l_1,m_1,l_2,m_2;l=0,m=0)$ is the Clebsch-Gordan(CG) coefficient for the decomposition of $|l, m\rangle$ in terms of $|l_1, m_2\rangle |l_2, m_2\rangle$, but the total angular momentum is fixed at $l=0$, and thus $m=0$. 

The following are the Jacobi coordinates for the baryon structure: we choose the coordinates set 1 as our reference.
\begin{itemize}
	\item{Coordinates Set 1}
	\begin{eqnarray}
	& \mathbf{x}_1 = \frac{1}{\sqrt{2}}({\mathbf r}_1 - {\mathbf r}_2), \qquad \mathbf{x}_2 = \frac{1}{\mu} \left( \frac{m_1 \mathbf{r}_1 + m_2 \mathbf{r}_2}{m_1 + m_2} - \mathbf{r}_3 \right) &\,,
	\nonumber
	\end{eqnarray}
	\item{Coordinates Set 2}
	\begin{eqnarray}
	& \mathbf{y}_1 = \frac{1}{\sqrt{2}}({\mathbf r}_1 - {\mathbf r}_3), \qquad \mathbf{y}_2 = \frac{1}{\mu} \left( \mathbf{r}_2 - \frac{m_1 \mathbf{r}_1 + m_2 \mathbf{r}_3}{m_1 + m_2} \right) &\,,
	\nonumber
	\end{eqnarray}
	\item{Coordinates Set 3}
	\begin{eqnarray}
	& \mathbf{z}_1 = \frac{1}{\sqrt{2}}({\mathbf r}_2 - {\mathbf r}_3), \qquad \mathbf{z}_2 = \frac{1}{\mu} \left( \frac{m_1 \mathbf{r}_2 + m_2 \mathbf{r}_3}{m_1 + m_2} - \mathbf{r}_1 \right) & \,,
	\nonumber
	\end{eqnarray}
\end{itemize}
where
\begin{eqnarray}
\mu &=& \left[ \frac{2 \left( m_1^2 + m_2^2 + m_1 m_2 \right)}{(m_1 + m_2)^2} \right]^{1/2} \,.	\nonumber
\end{eqnarray}
As in the mesons, the permutation symmetry of the spatial bases is determined by the angular momentum quantum numbers $l_1$ and $l_2$. As can be seen in the coordinates set 1, $\mathbf{x}_1$ is antisymmetric under the permutation (12), while $\mathbf{x}_2$ is symmetric. Therefore, the bases with even number of $l_1$ is symmetric, while the bases with odd number of $l_1$ is antisymmetric under the permutation (12).

In the case of the proton where all the constituent quarks are identical, the total wave function should be fully antisymmetric. Thus, we need to specify the symmetries for the permutations (13) and (23) as well. However, it is not clear in the coordinates set 1. Therefore, it is instructive to investigate the  method of constructing the spatial bases for the proton.

%%%%%%%%%%%%%%%%%%%%%%%%%%%%%%%%%%%%%%%%%%%%%%%%%%%%%%%%%%%%%%%%%%%%%%%%%%%%%%%%%%%%%%%%%%%%%%%
\subsection{Harmonic Oscillator Bases in Proton}
%%%%%%%%%%%%%%%%%%%%%%%%%%%%%%%%%%%%%%%%%%%%%%%%%%%%%%%%%%%%%%%%%%%%%%%%%%%%%%%%%%%%%%%%%%%%%%%
The proton is composed of three light quarks so that the symmetry property for the proton becomes complicated, compared to the other types of baryons. To satisfy the symmetry constraints on the proton structure, we can construct the spatial bases by using linear combinations of the harmonic oscillator bases. First, for the permutation group S$_3$, there are four possible Young tableaux as follows.
\begin{eqnarray}
\begin{tabular}{|c|c|c|}
  \cline{1-3}
  \,1\, & \,2\,  & \,3\, \\
  \cline{1-3}
\end{tabular} \,,\quad
\begin{tabular}{|c|c|}
  \hline
  \,1\,  & \,2\, \\
  \hline
  \,3\,  \\
  \cline{1-1}
\end{tabular} \,,\quad
\begin{tabular}{|c|c|}
  \hline
  \,1\,  & \,3\, \\
  \hline
  \,2\,  \\
  \cline{1-1}
\end{tabular} \,,\quad
\begin{tabular}{|c|}
	\cline{1-1}
	\,1\, \\
	\cline{1-1}
	\,2\, \\
	\cline{1-1}
	\,3\, \\
  \cline{1-1}
\end{tabular} \,.
\end{eqnarray}
There are correspondence relations between the Young tableaux and the harmonic oscillator bases such that, as the simplest examples,
\begin{eqnarray}
\begin{tabular}{|c|c|c|}
  \hline
  \,1\,  & \,2\,	& \,3\, \\
  \hline
\end{tabular} = \psi^{Spatial}_{[0,0,0,0]} (\mathbf{x}_1, \mathbf{x}_2)\,,
\end{eqnarray}
\begin{eqnarray}
\begin{tabular}{|c|c|}
  \hline
  \,1\,  & \,3\, \\
  \hline
  \,2\,  \\
  \cline{1-1}
\end{tabular} = \psi^{Spatial}_{[0,0,1,1]} (\mathbf{x}_1, \mathbf{x}_2)\,.
\end{eqnarray}
Using the follwing relation, we also obtain the basis corresponding to the Young tableau $\begin{small}
\begin{tabular}{|c|c|}
  \hline
  \,1\,  & \,2\, \\
  \hline
  \,3\,  \\
  \cline{1-1}
\end{tabular}
\end{small}$.

\begin{eqnarray}
\begin{tabular}{|c|c|}
  \hline
  \,1\,  & \,2\, \\
  \hline
  \,3\,  \\
  \cline{1-1}
\end{tabular}
&=&
\frac{2}{\sqrt{3}} \left[ (23) \,
\begin{tabular}{|c|c|}
  \hline
  \,1\,  & \,3\, \\
  \hline
  \,2\,  \\
  \cline{1-1}
\end{tabular}
- \frac{1}{2} \,
\begin{tabular}{|c|c|}
  \hline
  \,1\,  & \,3\, \\
  \hline
  \,2\,  \\
  \cline{1-1}
\end{tabular} \, \right]
\nonumber \\
&=&
\frac{2 \alpha}{\sqrt{3}} \left[ - \mathbf{y}_1 \cdot \mathbf{y}_2 - \frac{1}{2} \mathbf{x}_1 \cdot \mathbf{x}_2 \right] \exp \big[- a_1 x_1^2 - a_1 x_2^2 \big]
\nonumber \\
&=&
\frac{2 \alpha}{\sqrt{3}} \left[ - \frac{\sqrt{3}}{4} x_1^2 + \frac{\sqrt{3}}{4} x_2^2 \right] \exp \big[- a_1 x_1^2 - a_1 x_2^2 \big]
\nonumber \\
&=&
- \sqrt{\frac{1}{2}} \psi^{Spatial}_{[1,0,0,0]}(\mathbf{x}_1, \mathbf{x}_2) + \sqrt{\frac{1}{2}} \psi^{Spatial}_{[0,1,0,0]}(\mathbf{x}_1, \mathbf{x}_2) \,,
\nonumber \\
\end{eqnarray}
where $\alpha = - \sqrt{\frac{16}{3}} \left( \frac{2}{\pi} \right)^{\frac{3}{2}} a_1^{\frac{5}{2}}$ is the normalization factor, and the two variational parameters are taken to be the same $a_1=a_2$, which is due to the symmetry in the proton. For the Young tableau
$\begin{small}
\begin{tabular}{|c|}
  \hline
  \,1\,  \\
  \hline
  \,2\, \\
  \hline
  \,3\,  \\
  \cline{1-1}
\end{tabular}
\end{small}$\,, using the form of $\psi^{Spatial}_{[1,1,1,1]}$,
\begin{eqnarray}
\begin{tabular}{|c|}
  \hline
  \,1\,  \\
  \hline
  \,2\, \\
  \hline
  \,3\,  \\
  \cline{1-1}
\end{tabular}
&=&
\alpha \bigg\{ \psi^{Spatial}_{[1,1,1,1]}(\mathbf{x}_1, \mathbf{x}_2)
\nonumber \\
&& \hspace{1cm} + \psi^{Spatial}_{[1,1,1,1]}(\mathbf{y}_1, \mathbf{y}_2) + \psi^{Spatial}_{[1,1,1,1]}(\mathbf{z}_1, \mathbf{z}_2) \bigg\} \,,
\nonumber
\end{eqnarray}
where $\alpha$ is the normalization constant, and is manifestly antisymmetric under all the permutations (12), (13), and (23) as desired. After the transformations into the reference coordinates set $\{\mathbf{x}_1, \mathbf{x}_2\}$, and rearranging, it becomes:
\begin{eqnarray}
\begin{tabular}{|c|}
  \hline
  \,1\,  \\
  \hline
  \,2\, \\
  \hline
  \,3\,  \\
  \cline{1-1}
\end{tabular}
&=&
- \frac{\sqrt{3}}{4} \psi^{Spatial}_{[2,0,1,1]} (\mathbf{x}_1, \mathbf{x}_2) - \frac{\sqrt{3}}{4} \psi^{Spatial}_{[0,2,1,1]} (\mathbf{x}_1, \mathbf{x}_2)
\nonumber \\
&& + \sqrt{\frac{21}{40}} \psi^{Spatial}_{[1,1,1,1]} (\mathbf{x}_1, \mathbf{x}_2) - \sqrt{\frac{1}{10}} \psi^{Spatial}_{[0,0,3,3]} (\mathbf{x}_1, \mathbf{x}_2) \,.
\nonumber \\
\end{eqnarray}
In such a way above, it is possible to relate the harmonic oscillator bases to each of the Young tableaux for the permutation group S$_3$.

To fully construct the total wave function for the proton, it is necessary to construct the remaining parts of the wave function. For the color basis, it is always fully antisymmetric for the baryons. Thus, we focus on the spin and the isospin parts, and $\psi^{Spatial} \times \psi^{Isospin} \times \psi^{Spin}$ part should be fully symmetric. Since both the spin and the isospin bases are constructed by SU(2), we obtain the isospin-spin basis from the inner product of the flavor SU(2) and the spin SU(2). For the proton,
\begin{eqnarray}
\hspace{-0.7cm}&&\begin{tabular}{|c|c|}
  \hline
   \,\,\,\, & \,\,\,\, \\
  \hline
   \,\,\,\, \\
  \cline{1-1}
\end{tabular}_{\!\!\!I^{1/2}} \times \,\,\,
\begin{tabular}{|c|c|}
  \hline
   \,\,\,\, & \,\,\,\, \\
  \hline
   \,\,\,\, \\
  \cline{1-1}
\end{tabular}_{\!\!\!S^{1/2}}
\nonumber \\
\hspace{-0.7cm}&&=
\begin{tabular}{|c|c|c|}
  \hline
   \,\,\,\, & \,\,\,\,	& \,\,\,\, \\
  \cline{1-3}
\end{tabular}_{\,\,I^{1/2} S^{1/2}}
+ \,\,\,
\begin{tabular}{|c|c|}
  \hline
   \,\,\,\, & \,\,\,\, \\
  \hline
   \,\,\,\, \\
  \cline{1-1}
\end{tabular}_{\!\!\!\!I^{1/2} S^{1/2}}
+ \,\,\,
\begin{tabular}{|c|}
  \hline
   \,\,\,\, \\
  \hline
   \,\,\,\, \\
  \hline
   \,\,\,\, \\
  \cline{1-1}
\end{tabular}_{\,\,I^{1/2} S^{1/2}} \,,
\label{proton_decomp}
\end{eqnarray}
where the subscript $I^{1/2}$ ($S^{1/2}$) indicates the Young diagram corresponding to the isospin (spin) 1/2 state, and the Young diagrams on the right hand side in Eq.~(\ref{proton_decomp}) stand for SU(4), which correspond to the irreducible representations of the isospin-spin space for the proton. The irreducible representations on the right hand side in Eq.~(\ref{proton_decomp}) are spanned by the corresponding isospin-spin bases as follows. In terms of Young tableaux,
\begin{itemize}
\item{$\begin{tabular}{|c|c|}
  \hline
   \,\,\,\, & \,\,\,\, \\
  \hline
   \,\,\,\, \\
  \cline{1-1}
\end{tabular}_{\!\!\!\!I^{1/2} S^{1/2}}$}
\begin{eqnarray}
&&\hspace{-0.4cm}
\begin{tabular}{|c|c|}
  \hline
   \,1\, & \,2\, \\
  \hline
   \,3\, \\
  \cline{1-1}
\end{tabular}_{\!\!\!\!(IS)_{1}}
\nonumber \\
&=&
\frac{1}{\sqrt{2}}
\left(
\begin{tabular}{|c|c|}
  \hline
   \,1\, & \,2\, \\
  \hline
   \,3\, \\
  \cline{1-1}
\end{tabular}_{\!\!\!\!I_{1}} \times \,
\begin{tabular}{|c|c|}
  \hline
   \,1\, & \,2\, \\
  \hline
   \,3\, \\
  \cline{1-1}
\end{tabular}_{\!\!\!\!S_{1}} - \,
\begin{tabular}{|c|c|}
  \hline
   \,1\, & \,3\, \\
  \hline
   \,2\, \\
  \cline{1-1}
\end{tabular}_{\!\!\!\!I_{2}} \times \,
\begin{tabular}{|c|c|}
  \hline
   \,1\, & \,3\, \\
  \hline
   \,2\, \\
  \cline{1-1}
\end{tabular}_{\!\!\!\!S_{2}}
\right)\,,
\nonumber \\
&&\hspace{-0.4cm}
\begin{tabular}{|c|c|}
  \hline
   \,1\, & \,3\, \\
  \hline
   \,2\, \\
  \cline{1-1}
\end{tabular}_{\!\!\!\!(IS)_{2}}
\nonumber \\
&=&
-\frac{1}{\sqrt{2}}
\left(
\begin{tabular}{|c|c|}
  \hline
   \,1\, & \,2\, \\
  \hline
   \,3\, \\
  \cline{1-1}
\end{tabular}_{\!\!\!\!I_{1}} \times \,
\begin{tabular}{|c|c|}
  \hline
   \,1\, & \,3\, \\
  \hline
   \,2\, \\
  \cline{1-1}
\end{tabular}_{\!\!\!\!S_{2}} + \,
\begin{tabular}{|c|c|}
  \hline
   \,1\, & \,3\, \\
  \hline
   \,2\, \\
  \cline{1-1}
\end{tabular}_{\!\!\!\!I_{2}} \times \,
\begin{tabular}{|c|c|}
  \hline
   \,1\, & \,2\, \\
  \hline
   \,3\, \\
  \cline{1-1}
\end{tabular}_{\!\!\!\!S_{1}}
\right) \,.
\nonumber \\
\end{eqnarray}
%%%%%%%%%%%%%%%%%%%%%%%%%%%%%%%%%%%%%%%%%%%%%%%%%%
\item{$\begin{tabular}{|c|c|c|}
  \hline
   \,\,\,\, & \,\,\,\,	& \,\,\,\, \\
  \cline{1-3}
\end{tabular}_{\,\,I^{1/2} S^{1/2}}$}
\begin{eqnarray}
&&\hspace{-0.4cm}
\begin{tabular}{|c|c|c|}
  \hline
   \,1\, & \,2\,	& \,3\, \\
  \cline{1-3}
\end{tabular}_{\,\,(IS)_{3}}
\nonumber \\
&=&
\frac{1}{\sqrt{2}}
\left(
\begin{tabular}{|c|c|}
  \hline
   \,1\, & \,2\, \\
  \hline
   \,3\, \\
  \cline{1-1}
\end{tabular}_{\!\!\!\!I_{1}} \times \,
\begin{tabular}{|c|c|}
  \hline
   \,1\, & \,2\, \\
  \hline
   \,3\, \\
  \cline{1-1}
\end{tabular}_{\!\!\!\!S_{1}} + \,
\begin{tabular}{|c|c|}
  \hline
   \,1\, & \,3\, \\
  \hline
   \,2\, \\
  \cline{1-1}
\end{tabular}_{\!\!\!\!I_{2}} \times \,
\begin{tabular}{|c|c|}
  \hline
   \,1\, & \,3\, \\
  \hline
   \,2\, \\
  \cline{1-1}
\end{tabular}_{\!\!\!\!S_{2}}
\right) \,.
\nonumber \\
\end{eqnarray}
%%%%%%%%%%%%%%%%%%%%%%%%%%%%%%%%%%%%%%%%%%%%%%%%%%
\item{$\begin{tabular}{|c|}
  \hline
   \,\,\,\, \\
  \hline
   \,\,\,\, \\
  \hline
   \,\,\,\, \\
  \cline{1-1}
\end{tabular}_{\,\,I^{1/2} S^{1/2}}$}
\begin{eqnarray}
\hspace{0.7cm}
\begin{tabular}{|c|}
  \hline
   \,1\, \\
  \hline
   \,2\, \\
  \hline
   \,3\, \\
  \cline{1-1}
\end{tabular}_{\,\,(IS)_{4}}
\hspace{-0.7cm}=
\frac{1}{\sqrt{2}}
\left(
\begin{tabular}{|c|c|}
  \hline
   \,1\, & \,2\, \\
  \hline
   \,3\, \\
  \cline{1-1}
\end{tabular}_{\!\!\!\!I_{1}} \times \,
\begin{tabular}{|c|c|}
  \hline
   \,1\, & \,3\, \\
  \hline
   \,2\, \\
  \cline{1-1}
\end{tabular}_{\!\!\!\!S_{2}} - \,
\begin{tabular}{|c|c|}
  \hline
   \,1\, & \,3\, \\
  \hline
   \,2\, \\
  \cline{1-1}
\end{tabular}_{\!\!\!\!I_{2}} \times \,
\begin{tabular}{|c|c|}
  \hline
   \,1\, & \,2\, \\
  \hline
   \,3\, \\
  \cline{1-1}
\end{tabular}_{\!\!\!\!S_{1}}
\right) \,.
\nonumber \\
\end{eqnarray}
\end{itemize}
%%%%%%%%%%%%%%%%%%%%%%%%%%%%%%%%%%%%%%%%%%%%%%%%%%
where
$\begin{small}
\begin{tabular}{|c|c|}
  \hline
   \,1\, & \,2\, \\
  \hline
   \,3\, \\
  \cline{1-1}
\end{tabular}_{\!\!\!\!I_{1}(S_{1})}
\end{small}$ \!and \,\,
$\begin{small}
\begin{tabular}{|c|c|}
  \hline
   \,1\, & \,3\, \\
  \hline
   \,2\, \\
  \cline{1-1}
\end{tabular}_{\!\!\!\!I_{2}(S_{2})}
\end{small}$
are the isospin (spin) bases spanning the irreducible representation $\begin{small}
\begin{tabular}{|c|c|}
  \hline
   \,\,\,\, & \,\,\,\, \\
  \hline
   \,\,\,\, \\
  \cline{1-1}
\end{tabular}_{\!\!\!\!I^{1/2} (S^{1/2})}
\end{small}$ for the proton. On the other hand, in constructing the SU(4) irreducible representation of the type $\begin{small}
\begin{tabular}{|c|c|}
  \hline
   \,\,\,\, & \,\,\,\, \\
  \hline
   \,\,\,\, \\
  \cline{1-1}
\end{tabular}
\end{small}$, there are two more different methods of performing the inner product between any two of sptial, isospin, spin bases. However, these three methods of construction are equivalent.

We are now ready to construct the total wave function for the proton. Before proceeding, to avoid confusion, it is convenient to label the spatial bases  as in the isospin and spin bases as follows.
\begin{eqnarray}
\begin{tabular}{|c|c|}
  \hline
  \,1\,  & \,2\, \\
  \hline
  \,3\,  \\
  \cline{1-1}
\end{tabular}_{\!\!\!\!R_1} \,,\quad
\begin{tabular}{|c|c|}
  \hline
  \,1\,  & \,3\, \\
  \hline
  \,2\,  \\
  \cline{1-1}
\end{tabular}_{\!\!\!\!R_2} \,,\quad
\begin{tabular}{|c|c|c|}
  \cline{1-3}
  \,1\, & \,2\,  & \,3\, \\
  \cline{1-3}
\end{tabular}_{\,\,R_3} \,,\quad
\begin{tabular}{|c|}
	\cline{1-1}
	\,1\, \\
	\cline{1-1}
	\,2\, \\
	\cline{1-1}
	\,3\, \\
  \cline{1-1}
\end{tabular}_{\,\,R_4} \,.
\end{eqnarray}
There are three types of constructing the fully symmetric $\psi^{Spatial} \times \psi^{Isospin} \times \psi^{Spin}$.
\begin{eqnarray}
&&\begin{tabular}{|c|c|c|}
  \hline
   \,1\, & \,2\, & \,3\, \\
  \cline{1-3}
\end{tabular}_{\,\,(RIS)_1}
\nonumber \\
&=&
\frac{1}{\sqrt{2}}
\left(
\begin{tabular}{|c|c|}
  \hline
   \,1\, & \,2\, \\
  \hline
   \,3\, \\
  \cline{1-1}
\end{tabular}_{\!\!\!\!(IS)_{1}} \times \,
\begin{tabular}{|c|c|}
  \hline
   \,1\, & \,2\, \\
  \hline
   \,3\, \\
  \cline{1-1}
\end{tabular}_{\!\!\!\!R_{1}} + \,
\begin{tabular}{|c|c|}
  \hline
   \,1\, & \,3\, \\
  \hline
   \,2\, \\
  \cline{1-1}
\end{tabular}_{\!\!\!\!(IS)_{2}} \times \,
\begin{tabular}{|c|c|}
  \hline
   \,1\, & \,3\, \\
  \hline
   \,2\, \\
  \cline{1-1}
\end{tabular}_{\!\!\!\!R_{2}}
\right) \,,
%%%%%%%%%%%%%%%%%%%%%%%%%%%%%%%%%%%%
\nonumber \\
&&\begin{tabular}{|c|c|c|}
  \hline
   \,1\, & \,2\, & \,3\, \\
  \cline{1-3}
\end{tabular}_{\,\,(RIS)_2}
=
\,\,
\begin{tabular}{|c|c|c|}
  \hline
   \,1\, & \,2\, & \,3\, \\
  \cline{1-3}
\end{tabular}_{\,\,(IS)_3} \times \,\,\,
\begin{tabular}{|c|c|c|}
  \hline
   \,1\, & \,2\, & \,3\, \\
  \cline{1-3}
\end{tabular}_{\,\,R_3} \,,
%%%%%%%%%%%%%%%%%%%%%%%%%%%%%%%%%%%%
\nonumber \\
&&\begin{tabular}{|c|c|c|}
  \hline
   \,1\, & \,2\, & \,3\, \\
  \cline{1-3}
\end{tabular}_{\,\,(RIS)_3}
=
\,\,
\begin{tabular}{|c|}
  \hline
	\,1\, \\
  \cline{1-1}
  	\,2\, \\
  \cline{1-1}
  	\,3\, \\
  \cline{1-1}
\end{tabular}_{\,\,(IS)_4} \!\!\!\! \times \,\,\,
\begin{tabular}{|c|}
  \hline
	\,1\, \\
  \cline{1-1}
  	\,2\, \\
  \cline{1-1}
  	\,3\, \\
  \cline{1-1}
\end{tabular}_{\,\,R_4} \,.
\label{BasesType}
\end{eqnarray}

%%%%%%%%%%%%%%%%%%%%%%%%%%%%%%%%%%%%%%%%%%%%%%%%%%%%%%%%%%%%%%%%%%%%%%%%%%%%%%%%%%%%%%%%%%%%%%%
\subsection{Color, Spin Bases of Tetraquarks in a Baryon Structure}
%%%%%%%%%%%%%%%%%%%%%%%%%%%%%%%%%%%%%%%%%%%%%%%%%%%%%%%%%%%%%%%%%%%%%%%%%%%%%%%%%%%%%%%%%%%%%%%
As discussed in Section~\ref{BaryonlikeSec}, we regard the antidiquark as a point particle $Q_1$ which replaces ($\bar{Q}\bar{Q'}$). Labeling the quarks in the order of $q(1)$-$q'(2)$-$Q_1(3)$, for the total $S=1$ system, the spin bases can be constructed with the explicit spin numbers written in subscripts as follows.
\begin{eqnarray}
\left|1\right\rangle_1
&\equiv&
\left| \left[ q(1) q'(2) \right]_1 \left[ Q_1(3) \right]_0 \right\rangle \,,
\nonumber \\
\left|2\right\rangle_1
&\equiv&
\left| \left[ q(1) q'(2) \right]_1 \left[ Q_1(3) \right]_1 \right\rangle \,,
\nonumber \\
\left|3\right\rangle_1
&\equiv&
\left| \left[ q(1) q'(2) \right]_0 \left[ Q_1(3) \right]_1 \right\rangle \,.
\end{eqnarray}
Then the spin matrices with this bases set are obtained as follows.
\begin{eqnarray}
	\boldsymbol{\sigma}_1^{BL} \cdot \boldsymbol{\sigma}_2^{BL}
	&=&
	\left(
	\begin{array}{ccc}
	1	&	0	&	0	\\
	0	&	1	&	0	\\
	0	&	0	&	-3	\\
	\end{array}\right) \,,
	\nonumber \\
	\boldsymbol{\sigma}_1^{BL} \cdot \boldsymbol{\sigma}_3^{BL}
	&=&
	\left(
	\begin{array}{ccc}
	0	&		0		&	0	\\
	0	&		-2		&	-2\sqrt{2}	\\
	0	&	-2\sqrt{2}	&	0	\\
	\end{array}\right) \,,
	\nonumber \\
	\boldsymbol{\sigma}_2^{BL} \cdot \boldsymbol{\sigma}_3^{BL}
	&=&
	\left(
	\begin{array}{ccc}
	0	&		0		&	0	\\
	0	&		-2		&	2\sqrt{2}	\\
	0	&	2\sqrt{2}	&	0	\\
	\end{array}\right) \,,
\label{SpinMatBL}
\end{eqnarray}
where the superscript $BL$ indicates that they are the matrices in the baryon-like structure. On the other hand, the spin matrices in the tetraquark structure are as follows.
\begin{eqnarray}
	\boldsymbol{\sigma}_1 \cdot \boldsymbol{\sigma}_2
	&=&
	\left(
	\begin{array}{ccc}
	1	&	0	&	0	\\
	0	&	1	&	0	\\
	0	&	0	&	-3	\\
	\end{array}\right) \,,
	\nonumber \\
	\boldsymbol{\sigma}_1 \cdot \boldsymbol{\sigma}_3
	&=&
	\left(
	\begin{array}{ccc}
	0			&	\sqrt{2}	&	1	\\
	\sqrt{2}	&		-1		&	-\sqrt{2}	\\
	1			&	-\sqrt{2}	&	0	\\
	\end{array}\right) \,,
	\nonumber \\
	\boldsymbol{\sigma}_1 \cdot \boldsymbol{\sigma}_4
	&=&
	\left(
	\begin{array}{ccc}
	0			&	-\sqrt{2}	&	-1	\\
	-\sqrt{2}	&		-1		&	-\sqrt{2}	\\
	-1			&	-\sqrt{2}	&	0	\\
	\end{array}\right) \,,
	\nonumber \\
	\boldsymbol{\sigma}_2 \cdot \boldsymbol{\sigma}_3
	&=&
	\left(
	\begin{array}{ccc}
	0			&	\sqrt{2}	&	-1	\\
	\sqrt{2}	&		-1		&	\sqrt{2}	\\
	-1			&	\sqrt{2}	&	0	\\
	\end{array}\right) \,,
	\nonumber \\
	\boldsymbol{\sigma}_2 \cdot \boldsymbol{\sigma}_4
	&=&
	\left(
	\begin{array}{ccc}
	0			&	-\sqrt{2}	&	1	\\
	-\sqrt{2}	&		-1		&	\sqrt{2}	\\
	1			&	\sqrt{2}	&	0	\\
	\end{array}\right) \,,
	\nonumber \\
	\boldsymbol{\sigma}_3 \cdot \boldsymbol{\sigma}_4
	&=&
	\left(
	\begin{array}{ccc}
	-3	&	0	&	0	\\
	0	&	1	&	0	\\
	0	&	0	&	1	\\
	\end{array}\right) \,,
\label{SpinMatTetra}
\end{eqnarray}
with the spin bases
\begin{eqnarray}
\left|1\right\rangle
&\equiv&
\left| \left[ q(1) q'(2) \right]_1 \left[ \bar{Q}(3) \bar{Q}'(4) \right]_0 \right\rangle \,,
\nonumber \\
\left|2\right\rangle
&\equiv&
\left| \left[ q(1) q'(2) \right]_1 \left[ \bar{Q}(3) \bar{Q}'(4) \right]_1 \right\rangle \,,
\nonumber \\
\left|3\right\rangle
&\equiv&
\left| \left[ q(1) q'(2) \right]_0 \left[ \bar{Q}(3) \bar{Q}'(4) \right]_1 \right\rangle.
\end{eqnarray}
Comparing the spin matrices in Eq.~(\ref{SpinMatBL}) and Eq.~(\ref{SpinMatTetra}), one can find the following relations,
\begin{eqnarray}
\boldsymbol{\sigma}_1^{BL} \cdot \boldsymbol{\sigma}_3^{BL}
=
\boldsymbol{\sigma}_1 \cdot \boldsymbol{\sigma}_3 + \boldsymbol{\sigma}_1 \cdot \boldsymbol{\sigma}_4 \,,
\nonumber \\
\boldsymbol{\sigma}_2^{BL} \cdot \boldsymbol{\sigma}_3^{BL}
=
\boldsymbol{\sigma}_2 \cdot \boldsymbol{\sigma}_3 + \boldsymbol{\sigma}_2 \cdot \boldsymbol{\sigma}_4 \,.
\end{eqnarray}
The color bases in this baryon-like structure can be constructed with the explicit color states as follows.
\begin{eqnarray}
\psi^1_1
&\equiv&
\left[ q(1) q'(2) \right]^{\mathbf{6}} \left[ Q_1(3) \right]^{\bar{\mathbf{6}}} \,,
\nonumber \\
\psi^1_2
&\equiv&
\left[ q(1) q'(2) \right]^{\bar{\mathbf{3}}} \left[ Q_1(3) \right]^{\mathbf{3}} \,.
\end{eqnarray}
Then the color matrices with this bases set are obtained as follows.
\begin{eqnarray}
	{\boldsymbol{\lambda}^c_1}^{BL} {\boldsymbol{\lambda}^c_2}^{BL}
	&=&
	\left(
	\begin{array}{cc}
	\frac{4}{3}	&		0			\\
		0			&	-\frac{8}{3}	\\
	\end{array}\right) \,,
	\nonumber \\
	{\boldsymbol{\lambda}^c_1}^{BL} {\boldsymbol{\lambda}^c_3}^{BL}
	&=&
	\left(
	\begin{array}{cc}
	-\frac{20}{3}	&		0			\\
		0			&	-\frac{8}{3}	\\
	\end{array}\right) \,,
	\nonumber \\
	{\boldsymbol{\lambda}^c_2}^{BL} {\boldsymbol{\lambda}^c_3}^{BL}
	&=&
	\left(
	\begin{array}{cc}
	-\frac{20}{3}	&		0			\\
		0			&	-\frac{8}{3}	\\
	\end{array}\right) \,,
\label{ColorMatBL}
\end{eqnarray}
The color matrices in the tetraquark structure are as follows.
\begin{eqnarray}
	\boldsymbol{\lambda}^c_1 \boldsymbol{\lambda}^c_2
	&=&
	\left(
	\begin{array}{cc}
	\frac{4}{3}	&	0	\\
		0			&	-\frac{8}{3}	\\
	\end{array}\right) \,,
	\nonumber \\
	\boldsymbol{\lambda}^c_1 \boldsymbol{\lambda}^c_3
	&=&
	\left(
	\begin{array}{cc}
	-\frac{10}{3}	&	-2\sqrt{2}	\\
	-2\sqrt{2}		&	-\frac{4}{3}	\\
	\end{array}\right) \,,
	\nonumber \\
	\boldsymbol{\lambda}^c_1 \boldsymbol{\lambda}^c_4
	&=&
	\left(
	\begin{array}{cc}
	-\frac{10}{3}	&	2\sqrt{2}	\\
	2\sqrt{2}		&	-\frac{4}{3}	\\
	\end{array}\right) \,,
	\nonumber \\
	\boldsymbol{\lambda}^c_2 \boldsymbol{\lambda}^c_3
	&=&
	\left(
	\begin{array}{cc}
	-\frac{10}{3}	&	2\sqrt{2}	\\
	2\sqrt{2}		&	-\frac{4}{3}	\\
	\end{array}\right) \,,
	\nonumber \\
	\boldsymbol{\lambda}^c_2 \boldsymbol{\lambda}^c_4
	&=&
	\left(
	\begin{array}{cc}
	-\frac{10}{3}	&	-2\sqrt{2}	\\
	-2\sqrt{2}		&	-\frac{4}{3}	\\
	\end{array}\right) \,,
	\nonumber \\
	\boldsymbol{\lambda}^c_3 \boldsymbol{\lambda}^c_4
	&=&
	\left(
	\begin{array}{cc}
	\frac{4}{3}	&	0	\\
		0			&	-\frac{8}{3}	\\
	\end{array}\right) \,,
\label{ColorMatTetra}
\end{eqnarray}
with the color bases
\begin{eqnarray}
\psi^1_1
&\equiv&
\left[ q(1) q'(2) \right]^{\mathbf{6}} \left[ \bar{Q}(3) \bar{Q'}(4) \right]^{\bar{\mathbf{6}}} \,,
\nonumber \\
\psi^1_2
&\equiv&
\left[ q(1) q'(2) \right]^{\bar{\mathbf{3}}} \left[ \bar{Q}(3) \bar{Q'}(4) \right]^{\mathbf{3}} \,.
\end{eqnarray}
Likewise in the spin matrices, comparing Eq.~(\ref{ColorMatBL}) and Eq.~(\ref{ColorMatTetra}), one can also find similar relations as follows.
\begin{eqnarray}
{\boldsymbol{\lambda}^c_1}^{BL} {\boldsymbol{\lambda}^c_3}^{BL}
=
\boldsymbol{\lambda}^c_1 \boldsymbol{\lambda}^c_3 + \boldsymbol{\lambda}^c_1 \boldsymbol{\lambda}^c_4 \,,
\nonumber \\
{\boldsymbol{\lambda}^c_2}^{BL} {\boldsymbol{\lambda}^c_3}^{BL}
=
\boldsymbol{\lambda}^c_2 \boldsymbol{\lambda}^c_3 + \boldsymbol{\lambda}^c_2 \boldsymbol{\lambda}^c_4 \,.
\end{eqnarray}


\begin{thebibliography}{0}%
\makeatletter
\providecommand \@ifxundefined [1]{%
 \@ifx{#1\undefined}
}%
\providecommand \@ifnum [1]{%
 \ifnum #1\expandafter \@firstoftwo
 \else \expandafter \@secondoftwo
 \fi
}%
\providecommand \@ifx [1]{%
 \ifx #1\expandafter \@firstoftwo
 \else \expandafter \@secondoftwo
 \fi
}%
\providecommand \natexlab [1]{#1}%
\providecommand \enquote  [1]{``#1''}%
\providecommand \bibnamefont  [1]{#1}%
\providecommand \bibfnamefont [1]{#1}%
\providecommand \citenamefont [1]{#1}%
\providecommand \href@noop [0]{\@secondoftwo}%
\providecommand \href [0]{\begingroup \@sanitize@url \@href}%
\providecommand \@href[1]{\@@startlink{#1}\@@href}%
\providecommand \@@href[1]{\endgroup#1\@@endlink}%
\providecommand \@sanitize@url [0]{\catcode `\\12\catcode `\$12\catcode
  `\&12\catcode `\#12\catcode `\^12\catcode `\_12\catcode `\%12\relax}%
\providecommand \@@startlink[1]{}%
\providecommand \@@endlink[0]{}%
\providecommand \url  [0]{\begingroup\@sanitize@url \@url }%
\providecommand \@url [1]{\endgroup\@href {#1}{\urlprefix }}%
\providecommand \urlprefix  [0]{URL }%
\providecommand \Eprint [0]{\href }%
\providecommand \doibase [0]{http://dx.doi.org/}%
\providecommand \selectlanguage [0]{\@gobble}%
\providecommand \bibinfo  [0]{\@secondoftwo}%
\providecommand \bibfield  [0]{\@secondoftwo}%
\providecommand \translation [1]{[#1]}%
\providecommand \BibitemOpen [0]{}%
\providecommand \bibitemStop [0]{}%
\providecommand \bibitemNoStop [0]{.\EOS\space}%
\providecommand \EOS [0]{\spacefactor3000\relax}%
\providecommand \BibitemShut  [1]{\csname bibitem#1\endcsname}%
\let\auto@bib@innerbib\@empty
%</preamble>
\end{thebibliography}%


\begin{thebibliography}{99}

%\cite{Choi:2003ue}
\bibitem{Choi:2003ue} 
  S.~K.~Choi {\it et al.} [Belle Collaboration],
  %``Observation of a narrow charmonium - like state in exclusive B+- ---> K+- pi+ pi- J / psi decays,''
  Phys.\ Rev.\ Lett.\  {\bf 91}, 262001 (2003).
%  doi:10.1103/PhysRevLett.91.262001
%  [hep-ex/0309032].
  %%CITATION = doi:10.1103/PhysRevLett.91.262001;%%
  %1502 citations counted in INSPIRE as of 01 Aug 2018

%\cite{Ikeda:2016zwx}
\bibitem{Ikeda:2016zwx}
Y.~Ikeda \textit{et al.} [HAL QCD],
%``Fate of the Tetraquark Candidate $Z_c$(3900) from Lattice QCD,''
Phys. Rev. Lett. \textbf{117}, no.24, 242001 (2016)
doi:10.1103/PhysRevLett.117.242001
[arXiv:1602.03465 [hep-lat]].
%83 citations counted in INSPIRE as of 03 Jan 2021

%\cite{Nielsen:2009uh}
\bibitem{Nielsen:2009uh}
M.~Nielsen, F.~S.~Navarra and S.~H.~Lee,
%``New Charmonium States in QCD Sum Rules: A Concise Review,''
Phys. Rept. \textbf{497}, 41-83 (2010)
doi:10.1016/j.physrep.2010.07.005
[arXiv:0911.1958 [hep-ph]].
%274 citations counted in INSPIRE as of 03 Jan 2021

%\cite{Liu:2019zoy}
\bibitem{Liu:2019zoy}
Y.~R.~Liu, H.~X.~Chen, W.~Chen, X.~Liu and S.~L.~Zhu,
%``Pentaquark and Tetraquark states,''
Prog. Part. Nucl. Phys. \textbf{107}, 237-320 (2019)
doi:10.1016/j.ppnp.2019.04.003
[arXiv:1903.11976 [hep-ph]].
%189 citations counted in INSPIRE as of 03 Jan 2021

%\cite{DeRujula:1975qlm}
\bibitem{DeRujula:1975qlm}
A.~De Rujula, H.~Georgi and S.~L.~Glashow,
%``Hadron Masses in a Gauge Theory,''
Phys. Rev. D \textbf{12}, 147-162 (1975)
doi:10.1103/PhysRevD.12.147
%2683 citations counted in INSPIRE as of 03 Jan 2021



\bibitem{Bhaduri} 
  R. K. Bhaduri, L. E. Cohler, and Y. Nogami,
  %A Unified Potential for Mesons and Baryons,
  Nuovo Cim. \ A\  {\bf 65}, 376-390 (1981).
  
\bibitem{Silvestre:PRD1985} 
  B. Silvestre-Brac and C. Gignoux,
  %Study of Light Baryons in the Three-Quark-Cluster Model: Exact Calculations,
  Phys.\ Rev.\ D\ {\bf 32}, 743 (1985).
  
\bibitem{Silvestre:ZPC1993} 
  B. Silvestre-Brac and C. Semay,
  %Systematics of $L=0$ $q^2 \bar{q}^2$ systems,
  Z.\ Phys.\ C\  {\bf 57}, 273-282 (1993).
  
\bibitem{Silvestre:ZPC1994}
  C. Semay and B. Silvestre-Brac,
  %Diquonia and potential models,
  Z.\ Phys.\ C\  {\bf 61}, 271-275 (1994).
  
\bibitem{Silvestre:ZPC1986}
  C. Semay and B. Silvestre-Brac,
  %Diquonia and potential models,
  Z.\ Phys.\ C\  {\bf 30}, 457 (1986).
  
\bibitem{Vijande:PRD2009}
  J. Vijande, A. Valcarce, and N. Barnea,
  %Exotic meson-meson molecules and compact four-quark states,
  Phys.\ Rev.\ D\ {\bf 79}, 074010 (2009).
  
\bibitem{Brink:1998}
  D.M.~Brink and Fl.~Stancu,
  %Tetraquarks with heavy flavors,
  Phys.\ Rev.\ D\ {\bf 57}, 6778 (1998).
  
\bibitem{Janc:FewBody2004}
  D. Janc and M. Rosina,
  %The $T_{cc}=DD^*$ Molecular State,
  Few Body Syst.\ 35, 175 (2004).
    
\bibitem{Woosung:NPA2019} 
  Woosung Park, Sungsik Noh, and Su Houng Lee,
  %Masses of the doubly heavy tetraquarks in a constituent quark model,
  Nucl.\ Phys.\ A\  {\bf 983}, 1-19 (2019).
  %doi:10.1016/j.nuclphysa.2018.12.019
  %[arXiv:1809.05257 [nucl-th]]
     
\bibitem{Karliner:PRL2017}
  Marek Karliner, and Jonathan L. Rosner,
  %Discovery of the Doubly Charmed $\Xi_{cc}$ Baryon Implies a Stable $bb\bar{u}\bar{d}$ Tetraquark,
  Phys.\ Rev.\ Lett.\ {\bf 119}, 20, 202001 (2017)
  %doi:10.1103/PhysRevLett.119.202001
  %[arXiv:1707.07666 [hep-ph]].
\end{thebibliography}
\end{document}